# Intracavity Photon Statistics from Correlated Molecular Electronic Structure

Kurt R. Brorsen, Department of Chemistry, University of Missouri, Columbia, Missouri 65211, United States (brorsenk@missouri.edu)

## Abstract

Quantum optics characterizes light through photon correlations, motivating their connection to correlated *ab initio* molecular electronic structure. In this study, intracavity photon statistics of cavity-coupled molecules are computed from QED coupled-cluster (QED-CCSD-22) densities and validated against exact diagonalization of the Pauli–Fierz Hamiltonian. A change of coordinate origin or dipole convention displaces the ground state coherently, so second-order and higher cumulants of the cavity field are independent of these conventions even for molecular ions. The photon subsystem of QED Hartree–Fock is exactly coherent, so the invariant statistics describe fluctuations beyond this electron–photon product reference. The field fluctuations are super-Poissonian, and their degree of second-order coherence depends primarily on cavity frequency in the tested weak-coupling regime, falling from 18.5 to 4.9 for $H_2$ as the cavity is detuned from 5 to 20 eV. The third cumulant identifies non-Gaussian fluctuations and changes sign under reversal of the coupling direction.

In polaritonic chemistry, strong coupling between molecular transitions and the quantized field of an optical cavity alters chemical properties and reactivity, and molecular polaritons have been identified as a natural meeting point between chemistry and quantum optics.[1,2] Experimental data of molecular polaritons are obtained almost entirely through intensity-level observables such as transmission spectra, dispersion curves, and Rabi splittings.

In contrast, quantum optics characterizes light through its statistics such as the degree of second-order coherence $g^{(2)}(0)$, photon coincidence counting, Mandel statistics, and squeezing. Using model Hamiltonians, photon statistics of coupled light–matter systems have become a standard observable in theoretical quantum optics, with examples including photon correlations of few-level emitters in nanocavities and microcavities,[3,4] spatially and spectrally resolved $g^{(2)}$ maps of general nanophotonic systems,[5] multidimensional photon coincidence signals of polaritons,[6] semiclassical treatments of field correlations,[7] antibunching in single-molecule vibrational sum-frequency generation,[8] and photon-blockade and photon-statistics diagnostics of cavity-embedded quantum materials.[9–11] Intracavity photon observables have also been proposed as spectroscopic probes of strongly correlated matter, including an embedded $H_2$ molecule.[12] In all of these studies, however, the matter is described by a few-level, model-oscillator, or lattice Hamiltonian. Experimentally, photon statistics have already been measured for the emission of an organic molecule–cavity condensate.[13] Moreover, frequency- and time-resolved $g^{(2)}(\tau)$ measurements of single-molecule fluorescence have recently been reported.[14] These experiments probe an excited, emitting system and motivate molecular photon statistics without directly measuring the ground-state quantities considered here.

In contrast, *ab initio* polaritonic structure theory studies have treated the correlated many-electron molecule in full chemical detail while calculating simple photon field observables. QED Hartree–Fock (QED-HF) and QED coupled-cluster (QED-CC) methods,[15–18] which treat the molecule and the quantized cavity mode on equal footing, have been reviewed[19] and implemented at scale.[20] However, correlated ab initio photon-field studies have focused mainly on the mean photon number, as exemplified by a phaseless QED auxiliary-field quantum Monte Carlo study.[21]

We note two studies have moved toward more general photon statistics. Triana and Herrera computed intracavity photocount statistics and quadrature squeezing for a vibrational polariton, with the matter described by a single Morse oscillator and a model dipole function.[22] Tasci et al. used QED density-functional theory with a photon-many-body-dispersion functional, which maps the coupled system onto an effective quadratic-boson Hamiltonian.[23] They then computed ground-state photon-number fluctuations, $\langle \hat{a}^\dagger \hat{a}^\dagger \hat{a} \hat{a} \rangle$ (using their velocity-gauge mode operators), quadrature variances, and Mandel parameters for one-dimensional Ar chains, which was the first time such quantities had been calculated for chemical systems. We note that their photonic observables are evaluated in the velocity gauge, whereas the cumulants introduced below are displacement-invariant quantities of the length-gauge Hamiltonian. To our knowledge, this is the first correlated ab initio evaluation and exact-reference validation of displacement-invariant intracavity photon cumulants for neutral molecules and molecular ions.

In this Communication, we connect quantum-optical field statistics to molecular QED-CC by computing intracavity photon moments and validating against exact Pauli–Fierz diagonalization. We use the established QED-CCSD-22 and biorthogonal expectation-value machinery. Our primary contribution is the choice of convention-independent intracavity quantities, their exact-reference validation, and the resulting physical insights. The invariant statistics are specific to the molecule and describe fluctuations beyond the QED-HF product reference, thereby complementing the intensity-level quantities that are typically measured in molecular QED experiments. Raw and central moments coincide when the coherent amplitude vanishes, so invariant photon statistics are essential for polar and charged molecules. All results refer to the intracavity field, while emitted-light statistics require an open-system treatment.

The Pauli–Fierz (PF) Hamiltonian for a single cavity mode in the length gauge is

$$\hat{H} = \hat{H}_{\mathrm{el}} + \omega \hat{b}^{\dagger}\hat{b} - \sqrt{\omega/2}\,(\boldsymbol{\lambda}\cdot\hat{\boldsymbol{d}})\,(\hat{b}^{\dagger}+\hat{b}) + 1/2\,(\boldsymbol{\lambda}\cdot\hat{\boldsymbol{d}})^2, \quad (1)$$

where $\hat{H}_{\mathrm{el}}$ is the electronic Hamiltonian, $\hat{\boldsymbol{d}}$ is the molecular dipole operator, and $\omega$ and $\boldsymbol{\lambda}$ are the cavity frequency and coupling vector, respectively. The final term is the dipole self-energy (DSE), whose inclusion is essential to the invariance argument presented below. We use the coherent-state-transformed QED-HF reference,[15] and note that the coherent-state transformation is a coherent displacement of the photon mode. For the correlated wavefunction, we use QED-CCSD-22,[18,24] for which the cluster operator contains electronic single and double excitations, the pure photon excitations $\hat{b}^{\dagger}$ and $\hat{b}^{\dagger 2}$, and all mixed electron–photon products with up to two photons. Every statistic reported below is constructed from the ground-state expectation values of photon-operator strings (*i.e.*, the photon moments), which are defined as

$$M_{\hat{O}} = \langle \mathrm{HF}|(1+\hat{\Lambda})\,e^{-\hat{T}}\hat{O}\,e^{\hat{T}}|\mathrm{HF}\rangle, \quad (2)$$
$$\hat{O} \in \{\hat{b},\ \hat{b}^{\dagger},\ \hat{b}^{2},\ \hat{b}^{\dagger 2},\ \hat{b}^{\dagger}\hat{b},\ \hat{b}^{\dagger}\hat{b}^{2},\ \hat{b}^{\dagger 2}\hat{b},\ \hat{b}^{\dagger 2}\hat{b}^{2}\}$$

where $\hat{T}$ and $\hat{\Lambda}$ are the cluster and de-excitation operators, respectively. This same $\hat{\Lambda}$-dependent expectation-value structure underlies one-particle densities and analytic gradients in QED-CC.[25] All eight moment expressions were derived symbolically using SASQ[26] and evaluated in a locally developed code based on PySCF.[27] The explicit formulas are presented in the Zenodo deposit and discussed more in the supplemental information. We benchmarked the CC code against exact diagonalization of the PF Hamiltonian in the full configuration interaction (FCI) electronic space combined with a photon Fock space truncated at $n_{\max} = 20$. Increasing $n_{\max}$ from 12 to 20 changes $g_{\mathrm{c}}^{(2)}$, also defined below, by less than $2 \times 10^{-9}$. Finite biorthogonal CC does not guarantee positivity of $n_c$ or the fourth-order central moment. Both are positive in the tested data (Sec. S2), which is a numerical check rather than proof of a positive photon density operator.

The raw photon moments are defined once the field convention is specified. The dipole operator in Eq. (1) depends on the coordinate origin for charged systems, and on the convention chosen for $\hat{\boldsymbol{d}}$ (electronic-only versus total dipole) even for neutral systems. Naively computed photon numbers and values of the degree of second-order coherence $g^{(2)}(0) = \langle \hat{b}^{\dagger}\hat{b}^{\dagger}\hat{b}\hat{b}\rangle / \langle \hat{b}^{\dagger}\hat{b}\rangle^{2}$ inherit this arbitrariness. Schäfer et al.[28] showed that an origin shift coherently displaces the photon subsystem and replaced the $\hat{b}^{\dagger}\hat{b}$ occupation by an operator-valued occupation built from field observables. Haugland et al.[15] formulated QED-HF in the coherent-state basis, wrote the coupling and self-energy in terms of dipole fluctuations, and established the origin invariance of the QED-HF energy. Coherent-state relaxation likewise restores the origin invariance of correlated QED energies for charged systems,[29] and related coherent-displacement constructions appear at the orbital level[30] and in a recent analysis of the coherent-state transformation in QED-CC.[31] Most recently, Zhang and Liu[32] identified an origin-tunable Poissonian photon-number distribution for cavity-coupled ions and analyzed its origin dependence in terms of Fock-space basis convergence. These results motivate intracavity statistics that are independent of the coherent displacement. With the full DSE, shifting the dipole by any c-number $c$ gives

$$\hat{H}(\boldsymbol{\lambda}\cdot\hat{\boldsymbol{d}}+c) \;=\; \hat{D}(\alpha)\,\hat{H}(\boldsymbol{\lambda}\cdot\hat{\boldsymbol{d}})\,\hat{D}^{\dagger}(\alpha), \qquad \hat{D}(\alpha)=e^{\alpha(\hat{b}^{\dagger}-\hat{b})}, \quad \alpha=\frac{c}{\sqrt{2\omega}}. \tag{3}$$

---

The identity follows because the full DSE completes the square in the photon coordinate. Under any such shift, the exact ground state therefore changes only by a coherent displacement. In quantum optics language, a coherent displacement changes only the coherent amplitude $\langle\hat{b}\rangle$, which is the first cumulant of the field. All central moments of the fluctuation operator, and therefore all second- and higher-order cumulants, are origin- and displacement-invariant. The observables of this work, the incoherent photon number, $n_{\mathrm{c}}$, and the degree of second-order coherence of the field fluctuations, $g_{\mathrm{c}}^{(2)}$, are defined as

$$n_{\mathrm{c}}=\langle\delta\hat{b}^{\dagger}\delta\hat{b}\rangle, \qquad g_{\mathrm{c}}^{(2)}=\frac{\langle\delta\hat{b}^{\dagger}\delta\hat{b}^{\dagger}\delta\hat{b}\,\delta\hat{b}\rangle}{{n_{\mathrm{c}}}^{2}}, \qquad \delta\hat{b}=\hat{b}-\langle\hat{b}\rangle, \tag{4}$$

---

where $\delta\hat{b}$ is the fluctuation operator, the subscript c denotes central moments, and in the CC case each operator is shifted by its biorthogonal mean.

In the standard decomposition of quantum optics, the intracavity intensity separates into coherent and incoherent parts as $\langle\hat{b}^{\dagger}\hat{b}\rangle=|\langle\hat{b}\rangle|^{2}+n_{\mathrm{c}}$, with the biorthogonal product $\langle b^{\dagger}\rangle\langle b\rangle$ replacing $|\langle b\rangle|^{2}$ at finite CC truncation, so $n_{c}$ is the incoherent photon number and $g_{\mathrm{c}}^{(2)}$ is the degree of second-order coherence of the incoherent component. We note that $n_{\mathrm{c}}$ is distinct from the photon-number variance $\langle(\Delta\hat{n})^{2}\rangle$ that enters the Mandel parameter, and the invariant Mandel parameters and quadrature variances are given in the supplementary material. Reporting photon numbers in the coherent-state frame is not equivalent, because that frame is tied to the mean-field dipole and its photon number differs from $n_{\mathrm{c}}$ for a correlated state. The operator-valued occupation of Ref. 28 addresses a different question from the central moments of the canonical mode used here.

This coherent-displacement invariance is distinct from gauge invariance. The Power–Zienau–Woolley transformation mixes matter and field operators, so comparing the same physical observable between gauges requires consistent transformation of the Hamiltonian, state, and observable.[33,34] Mean subtraction does not remove this operator-valued change. The present calculations establish length-gauge intracavity statistics, without testing numerical equivalence between gauges at finite truncation.

We find that rigid translations of $H_2$ and $HeH^+$ spanning coherent shifts $z\in[-1.08,1.18]$ change the raw $g^{(2)}(0)$ from 1.00 to 15.2, while $g_{\mathrm{c}}^{(2)}$ is constant to a relative spread of $2.6\times10^{-12}$ for $H_2$ and $5.2\times10^{-10}$ for $HeH^+$. Raw intracavity $g^{(2)}(0)$ values without a dipole convention are therefore not meaningful. For a neutral molecule the total dipole does not depend on the coordinate origin, so raw moments computed with the total-dipole convention are well defined. However, they differ from the invariant cumulants through the static coherent amplitude $\langle\hat{b}\rangle$, whose subtraction alone does not determine detected correlations.

The unnormalized central moments vanish identically at the mean-field level. The QED-HF ground state in the coherent-state frame is $|\mathrm{HF}\rangle\otimes|0\rangle$,[15] so in the lab frame its photon subsystem is exactly a coherent state and its fluctuation statistics are those of the vacuum, with

$n_c \equiv 0$, $g_c^{(2)}$ undefined, and raw $g^{(2)}(0) = 1$ when the coherent amplitude is nonzero. Thus, the photon fluctuation statistics describe fluctuations beyond the coherent-state electron–photon product reference. We note that they do not separately measure electron–electron correlation. Electronic correlation changes the excitation energies, dipole matrix elements, and mixed amplitudes entering these moments. Increasing electronic excitation rank need not change $g_c^{(2)}$ monotonically.

Matched-basis QED-FCI comparisons isolate coupled-cluster truncation error from electronic basis incompleteness. STO-3G/QED-FCI comparisons have been used in previous QED-CC and quantum-algorithm studies.[15,24,35] To validate our method, we first performed calculations on two-electron systems for which CCSD is electronically exact, so any deviation from exact diagonalization isolates the truncation of the photon excitation rank of the cluster operator. For $H_2$ (6-31G, $R = 0.74$ Å) and $HeH^+$ (STO-3G, $R = 0.775$ Å) with $\boldsymbol{\lambda}$ along the bond, the ground-state energies agree with exact diagonalization to better than $10^{-12}$ Ha for couplings $\lambda \leq 0.01$ a.u. and degrade to $\sim 5 \times 10^{-8}$ Ha at $\lambda = 0.1$ a.u. Figure 1 shows the relative error of the CC $g_c^{(2)}$ against exact diagonalization as a function of coupling strength. The invariant $g_c^{(2)}$ agrees to $\approx 4 \times 10^{-5}$ relative at $\lambda = 0.005$ a.u., and the error grows as $\lambda^2$ (fitted exponent 1.99), reaching $1.7\%$ at $\lambda = 0.1$ a.u.

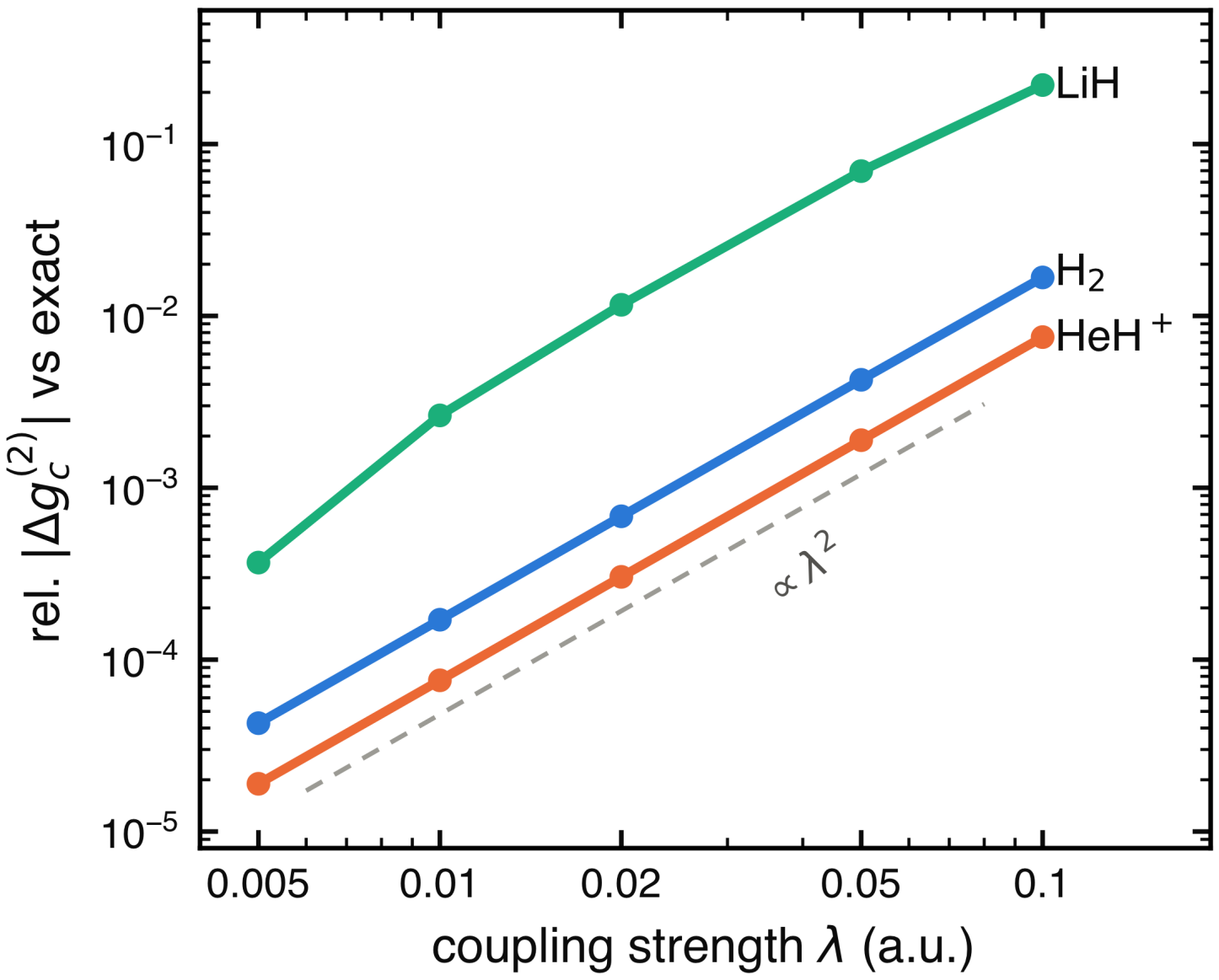

**Figure 1**: *Relative error of $g_c^{(2)}$ against exact diagonalization versus $\lambda$ at $\omega = 10$ eV. For the two-electron systems ($H_2$, $HeH^+$), CCSD is electronically exact and the error is pure photon-rank truncation, scaling as $\lambda^2$ (gray guide). For LiH, the error sits on an electronic floor already at $\lambda = 0.005$ a.u. and reaches $22\%$ at $\lambda = 0.1$ a.u.*

LiH/STO-3G ($R = 1.595$ Å) is the smallest system for which CCSD is not electronically exact, and its $g_c^{(2)}$ error is an order of magnitude above the two-electron curves already at the weakest coupling ($3.7 \times 10^{-4}$ at $\lambda = 0.005$ a.u.), while the energy error shows a $\lambda$-independent electronic floor of $1.05 \times 10^{-5}$ Ha. The $g_c^{(2)}$ error reaches $22\%$ at $\lambda = 0.1$ a.u., where CC underestimates the coherence ($g_c^{(2)}$ of 12.28 versus the exact 15.76). The larger LiH error indicates substantial contributions from truncated electronic and mixed electron–photon excitation ranks. The larger-basis LiH/6-31G benchmark gives a 28.3% error at $\lambda = 0.1$ a.u. (Sec. S11, Table S18). At $\lambda =$ $0.02$ a.u., the errors are 1.16% in STO-3G and 1.76% in 6-31G. Small total-energy errors do not bound the relative error in $g_c^{(2)}$, whose fourth-order numerator probes small two-photon components and is normalized by $n_c^2$. The 0.1 a.u. endpoint tests the limits of the approximation, while the molecular frequency scan in Fig. 5 uses 0.02 a.u.

The lab-frame examples in Table S2 use $\lambda = 0.05$ a.u., $\omega = 10$ eV, and the electronic-position dipole convention, and require support up to $n \approx 6$–8 even though $n_c$ itself is small. Truncating the FCI reference at $n_{\max} = 2$ gives an incorrect estimate of $g_c^{(2)}$ by $+11\%$ for $H_2$ and by a factor of 12 for LiH. QED-CCSD-22 is distinct from a hard two-photon cutoff because the exponential ansatz still populates high photon Fock states through disconnected powers of the photon cluster operators. The required cutoff depends on coupling, frequency, coherent displacement, and requested accuracy, so the chosen statistic must be converged in its chosen frame.

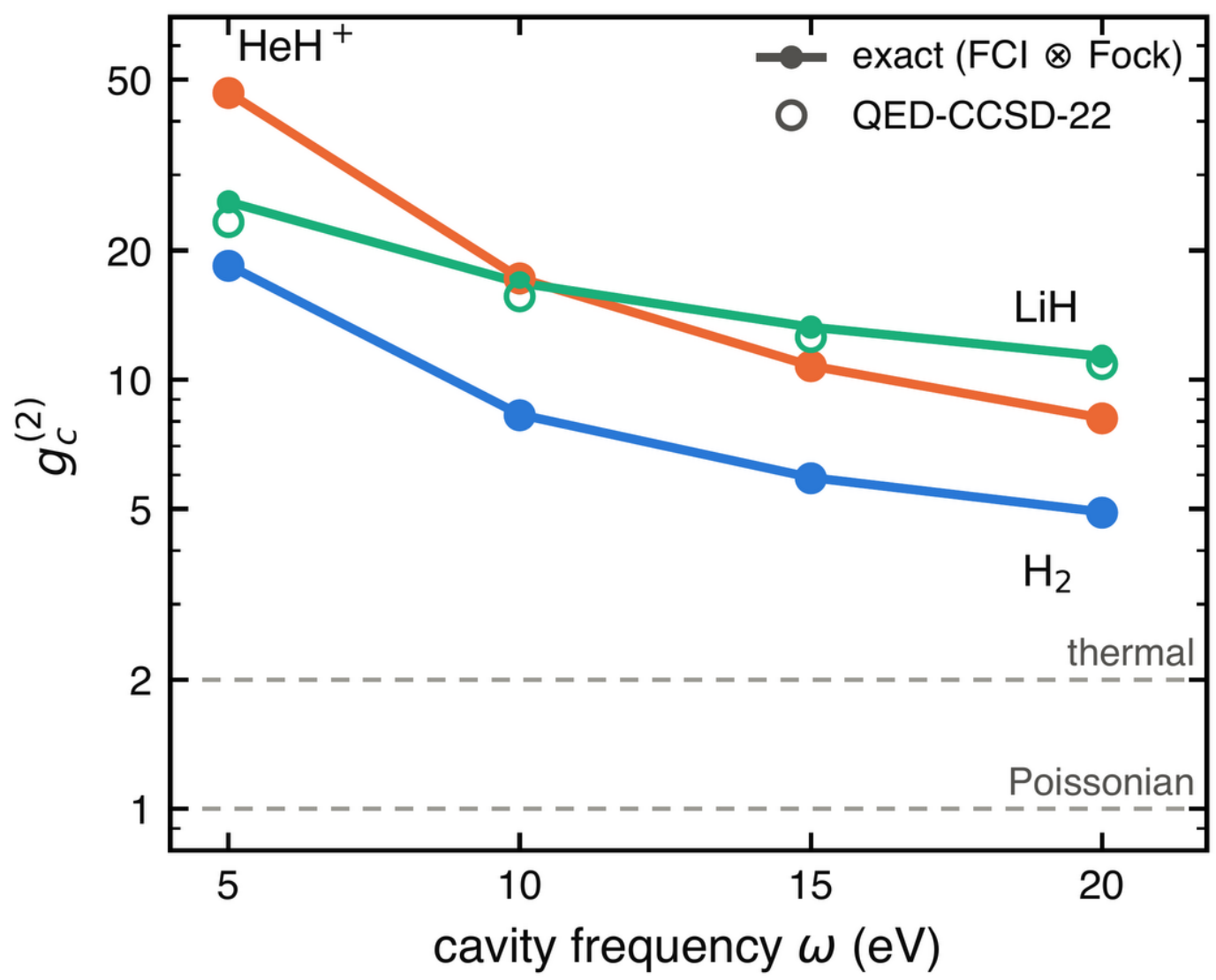

**Figure 2**: *Cavity-frequency dependence of the degree of second-order coherence of the field fluctuations, $g_c^{(2)}$, at $\lambda = 0.05$ a.u. for $H_2$/6-31G, $HeH^+$/STO-3G, and LiH/STO-3G. Solid lines with filled markers denote exact Pauli–Fierz diagonalization, and open markers denote QED-CCSD-22. Dashed horizontal lines mark Poissonian ($g^{(2)} = 1$) and thermal ($g^{(2)} = 2$) statistics. The coherence $g_c^{(2)}$ grows monotonically at low cavity frequency.*

Figure 2 shows the cavity-frequency dependence of $g_c^{(2)}$ for $H_2$, $HeH^+$, and LiH. The intracavity field fluctuations are strongly super-Poissonian, that is, photon bunched, with $g_c^{(2)} \approx 5$–$47$ across the systems and frequencies studied, and the cavity frequency is the dominant tuning parameter. Detuning the cavity from $\omega = 5$ to 20 eV lowers $g_c^{(2)}$ from 18.5 to 4.9 for $H_2$, from 46.6 to 8.1 for $HeH^+$, and from 26.0 to 11.3 for LiH, so the effect is strongest at the low-frequency end of the window.

The incoherent photon number $n_c$ is complementary to these results (Fig. 3). A fine frequency scan for $H_2$ ($0.25$ eV steps) resolves a broad maximum of $n_c$, $35\%$ above its $5$ eV value, at $\omega = 15.5$ eV, and the FCI $\sigma \rightarrow \sigma^*$ excitation in the same electron basis set lies at $15.31$ eV. In contrast, $g_c^{(2)}$ passes through the resonance without structure, and the curvature of $\ln\left(g_c^{(2)}\right)$ is an order of magnitude smaller in the resonance window than at the low-frequency end of the scan.

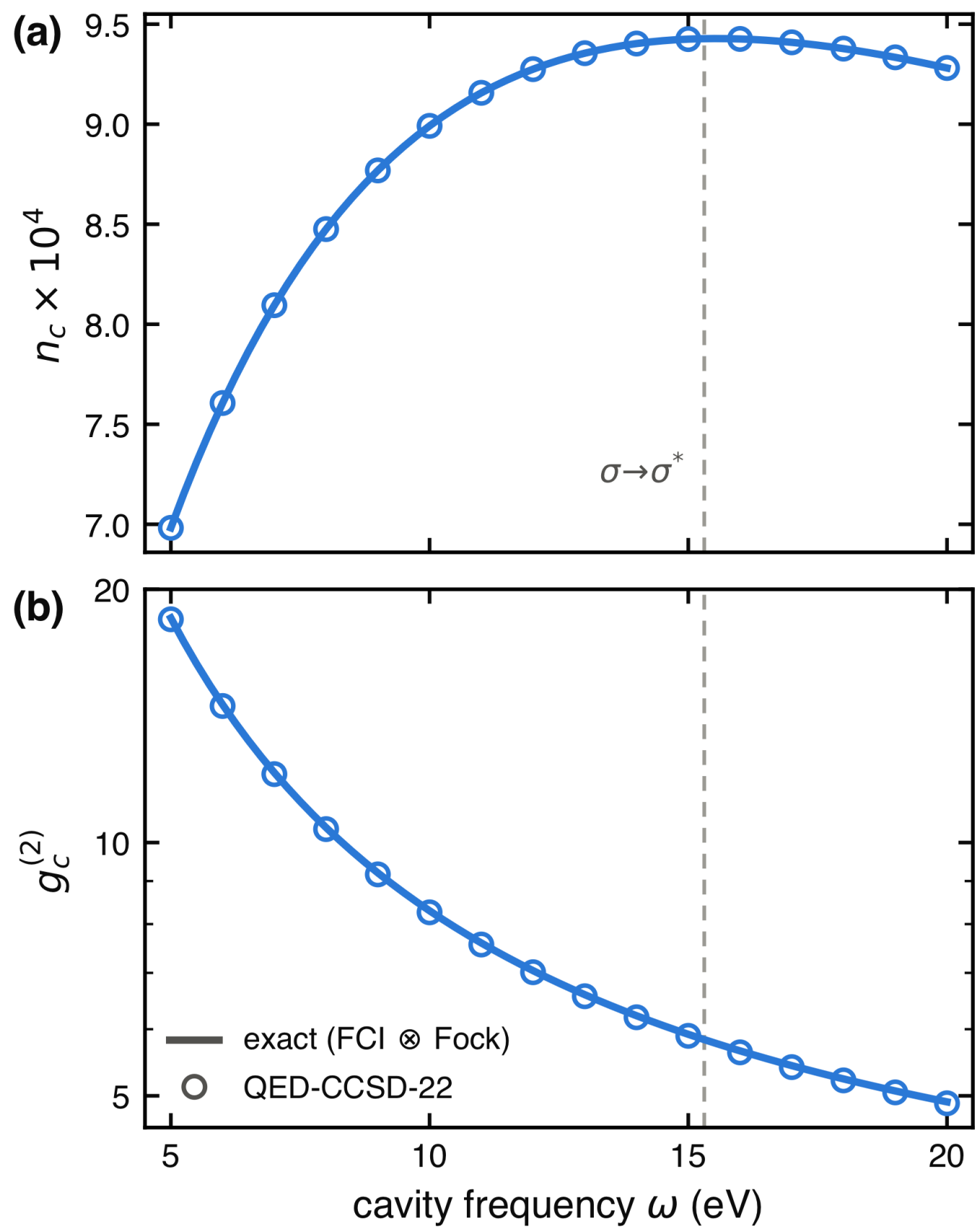

*Figure 3: Resonance behavior of the photon statistics. Fine cavity-frequency scan for $H_2$/6-31G at $\lambda = 0.05$ a.u. (0.25 eV steps): exact diagonalization (solid) and QED-CCSD-22 (open markers, every fourth point). (a) The incoherent photon number $n_c$ shows a broad maximum at $\omega = 15.5$ eV, adjacent to the FCI $\sigma \to \sigma^*$ excitation at 15.31 eV (dashed vertical line). (b) $g_c^{(2)}$ decays through the resonance without structure.*

To further investigate these trends, we compared them with a closed-form two-level model. For a single bright transition at energy $\Delta E$ with transition dipole $d$, leading-order perturbation theory gives $n_c \propto \lambda^2 d^2 \omega/(\Delta E + \omega)^2$ and $g_{\mathrm{c}}^{(2)} = [(\Delta E + \omega)/\omega]^2$. Ground-state dressing involves sums such as $\Delta E + \omega$, so no denominator becomes small at $\omega = \Delta E$. The $2\omega$ two-photon denominator is cancelled by the $\omega$ growth of the product of the two coupling matrix elements. With $\Delta E$ set to the $\sigma \to \sigma^*$ energy and the scale fixed at a single point, the model reproduces the exact $n_{\mathrm{c}}$ scan to within $0.5\%$ at all 61 frequencies, places the maximum exactly at $\omega = \Delta E$, and yields the $35\%$ enhancement. The exact $g_{\mathrm{c}}^{(2)}$ exceeds the two-level value by $12\%$ at $\omega = 5$ eV, growing monotonically to $58\%$ at 20 eV. This excess reflects dipole-fluctuation pathways beyond a single transition and motivates a multistate treatment.

The origin dependence is most severe for the charged systems. QED energies computed without the coherent-state transformation have been shown to vary with the coordinate origin for the hydrogen fluoride cation,[29] and the photon-number distribution of a cavity-coupled ion is origin-tunable.[32] The invariant cumulants in this study remove this ambiguity. Equation (3) does not assume a neutral system, so photon statistics of molecular ions are well-defined, origin-invariant objects. To the best of our knowledge, the $HeH^+$ frequency scan of Fig. 2 is the first well-defined, origin-invariant photon statistics reported for a molecular ion using *ab initio* QED methods.

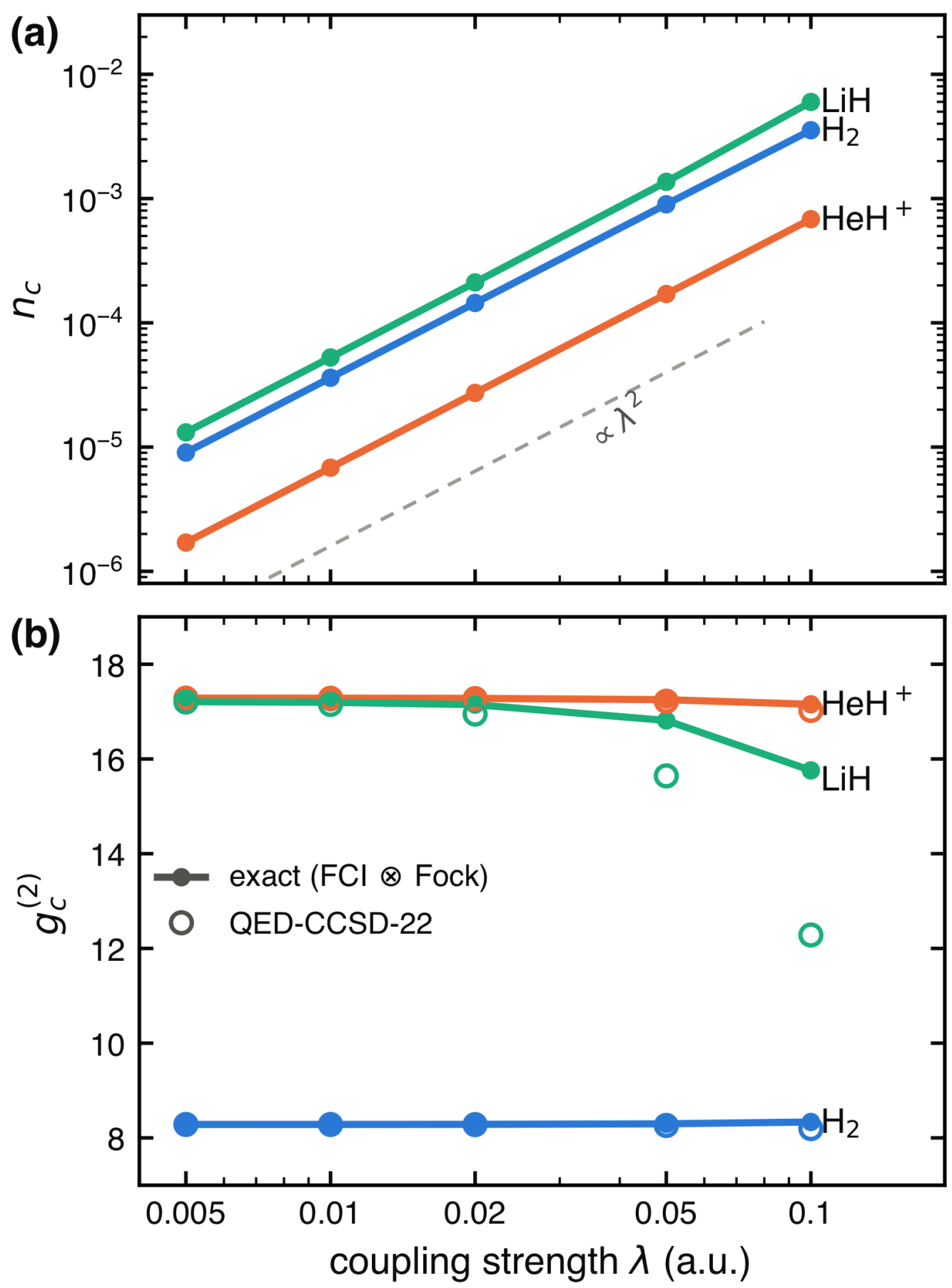

**Figure 4**: *Coupling-strength dependence at $\omega = 10$ eV. (a) Incoherent photon number $n_c$ versus $\lambda$ (exact diagonalization; QED-CCSD-22 values are indistinguishable on this scale); the gray guide indicates $n_c \propto \lambda^2$. (b) $g_c^{(2)}$ versus $\lambda$: exact (solid, filled) and QED-CCSD-22 (open markers). The exact $g_c^{(2)}$ is nearly flat over a twentyfold range of $\lambda$, while the CC value for LiH drifts below the exact result as coupling grows, reaching $-22\%$ at $\lambda = 0.1$ a.u.*

The coupling strength only weakly changes the statistics (Fig. 4). Over the range $\lambda = 0.005$–$0.1$ a.u., the exact $g_c^{(2)}$ at fixed $\omega = 10$ eV changes by $0.63\%$ for $H_2$ and by $0.76\%$ for $HeH^+$. For LiH, it changes by $2.4\%$ up to $\lambda = 0.05$ a.u. and by $9\%$ over the full range. The coupling instead determines the size of the fluctuations, with $n_c \propto \lambda^2$ (fitted exponent 1.99) ranging from $10^{-6}$ to $6 \times 10^{-3}$ photons across the grid. At weak coupling, one- and two-photon amplitudes first appear at orders $\lambda$ and $\lambda^2$, giving $n_c = \lambda^2 A(\omega) + O(\lambda^4)$ and $G_c^{(2)} = \lambda^4 B(\omega) + O(\lambda^6)$, where $G_c^{(2)} = \langle \delta b^\dagger \delta b^\dagger \delta b \delta b \rangle$. The leading coupling powers cancel in $g_c^{(2)} = B/A^2 + O(\lambda^2)$, while the frequency-dependent weights of the virtual transitions remain. In the weak-coupling limit $g_c^{(2)}$ reduces to a ratio of dipole-fluctuation response functions of the isolated molecule, with the 9% LiH drift indicating departures at larger coupling.

Enlarging the $H_2$ basis set from STO-3G to 6-31G lowers $g_c^{(2)}$ at $\omega = 10$ eV and $\lambda = 0.01$ a.u. from 15.25 to 8.28. Predictions for molecules beyond two electrons are therefore made using the aug-cc-pVDZ basis set, with the small-basis STO-3G calculations serving to validate against exact diagonalization. Small coupling suppresses coupling-dependent truncation errors, but cannot eliminate errors in the electronic response that remain in the weak-coupling ratio.

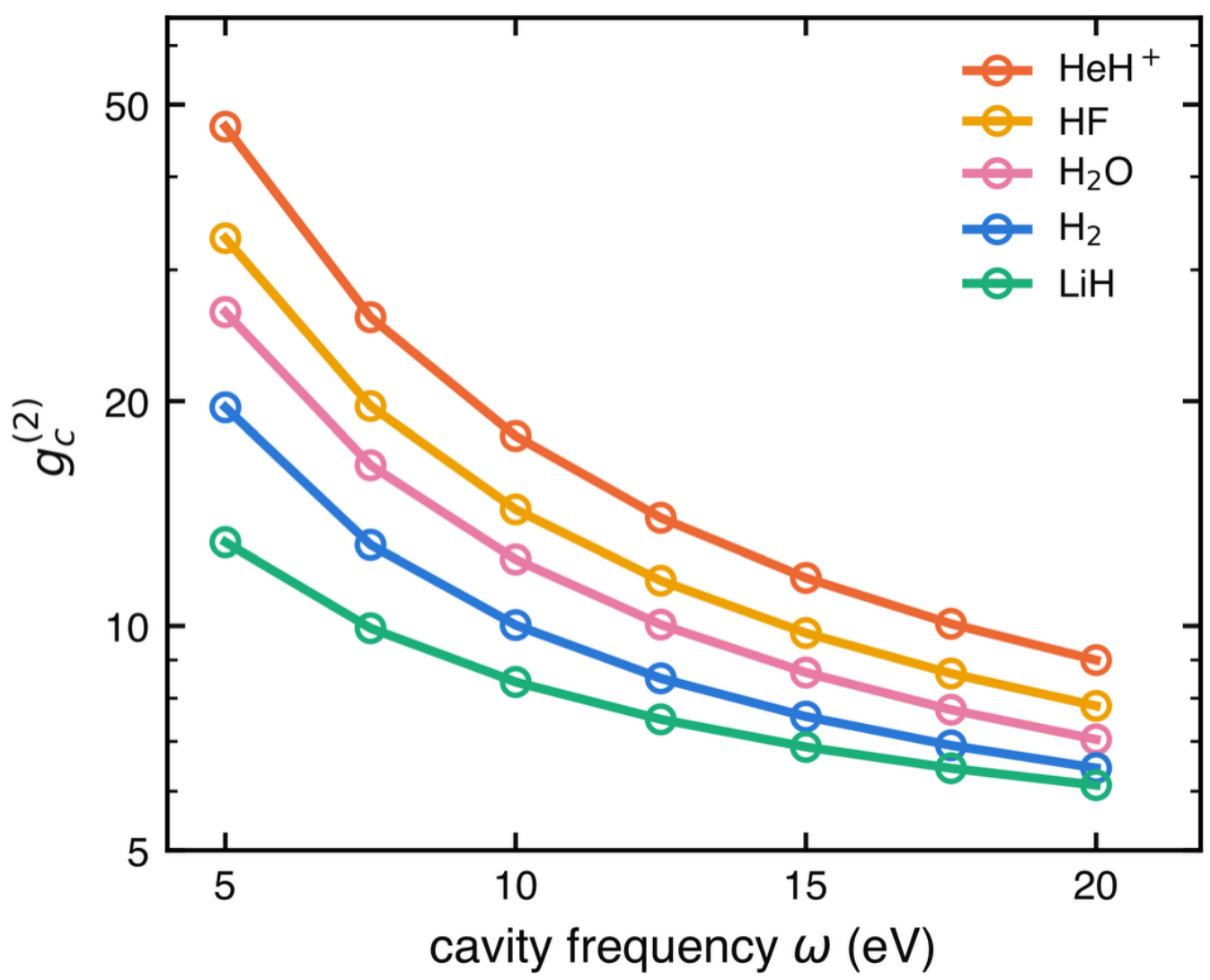

**Figure 5**: *$g_c^{(2)}$ versus cavity frequency $\omega$ from QED-CCSD-22 at the aug-cc-pVDZ level ($\lambda = 0.02$ a.u. along the molecular axis; $C_{2v}$ axis for $H_2O$) for $H_2$, HF, $H_2O$, $HeH^+$, and LiH; open markers denote CC values. For HF, $H_2O$, and LiH, no exact reference is feasible at this basis, while the two-electron systems are validated directly. The five-system ordering is fixed across the window, and the separation is largest at its low-frequency end.*

Figure 5 plots $g_c^{(2)}$ versus cavity frequency $\omega$ for $H_2$, HF, $H_2O$, $HeH^+$, and LiH using the aug-cc-pVDZ basis set ($\lambda = 0.02$ a.u., $\omega = 5$–$20$ eV). We find that low-frequency super-Poissonian statistics occur in all five systems, with $g_c^{(2)}$ falling monotonically with $\omega$ for every system, from 33.1 to 7.8 for HF, from 26.3 to 7.0 for $H_2O$, and from 19.6 to 6.4 for $H_2$. The ordering $\mathrm{HeH}^+ > \mathrm{HF} > \mathrm{H_2O} > \mathrm{H_2} > \mathrm{LiH}$ is fixed at every frequency in the window. All five calculations share the same coupling strength, so the differences encode correlation-driven dipole-fluctuation structure specific to each molecule. Therefore, at specified geometry, orientation, and cavity settings, the statistics constitute a molecular fingerprint.

The difference among molecules is strongest at low frequency, where the curves span a factor of 3.6 at $\omega = 5$ eV compared with 1.5 at 20 eV. Moreover, the ordering is not set by the lowest excitation energy alone. Reproducing the HF value at $\omega = 5$ eV requires an effective transition energy near 24 eV in the two-level model, well above the lowest excitation of HF that couples to the cavity polarization, and $H_2$ lies below HF and $H_2O$ despite its higher first excitation. In the two-level expression, increasing the excitation energy increases $g_c^{(2)}$, while the overall transition-dipole magnitude cancels. For several states, successive virtual transitions depend on the relative transition strengths and intermediate-state dipole couplings, which electronic correlation also changes. Together with the $H_2$ discrepancy above, these results support a multistate interpretation of the ordering without assigning each molecule to a single transition.

Enlarging aug-cc-pVDZ to aug-cc-pVTZ shifts $g_{\mathrm{c}}^{(2)}$ by at most $2\%$ across the $\omega = 5–20$ eV checks for $H_2$, HF, and $H_2O$, whereas removing the diffuse functions (cc-pVDZ) shifts it by $5–26\%$. Diffuse augmentation is therefore essential for a correct description of the invariant statistics. Adding a second diffuse shell (d-aug-cc-pVDZ) moves the checks by at most $1.2\%$ for $H_2$ and $2.1\%$ for HF, but by up to $5.1\%$ for $H_2O$ at $\omega = 5$ eV (supplementary material). However, the five-system ordering of Fig. 5 remains unchanged under these shifts at every frequency checked.

$\lambda$-independence also persists at the aug-cc-pVDZ level, where over the tenfold range $\lambda = 0.005–0.05$ a.u., $g_{\mathrm{c}}^{(2)}$ at $\omega = 10$ eV changes by $1.7\%$ for $H_2$, $1.6\%$ for HF, and $2.0\%$ for $H_2O$. For $H_2$ and $HeH^+$ exact diagonalization is possible using aug-cc-pVDZ and reproduces the QED-CCSD-22 $g_{\mathrm{c}}^{(2)}$ to within $0.24\%$ and $0.06\%$, respectively, at every frequency of Fig. 5. These comparisons and the LiH results support small errors for the tested systems at the coupling used in Fig. 5, but do not bound the electronic truncation error for HF or $H_2O$.

Central moments coincide with cumulants through third order, so $\kappa_{3,\mathrm{c}} = \langle \delta\hat{b}^{\dagger}\delta\hat{b}\,\delta\hat{b}\rangle$ is a third-order joint field cumulant. Like the second-order moments, its working expression contains no coherent shift, so it is origin-invariant, and vanishes identically at QED-HF. Moreover, it vanishes for every Gaussian state, so a nonzero value is due to non-Gaussian photon fluctuations, which no effective quadratic-boson description can reproduce.

For centrosymmetric molecules, the third-order moment also obeys a parity selection rule. With $\boldsymbol{\lambda}$ along the symmetry axis, combined electronic inversion and $\hat{b} \to -\hat{b}$ forces every odd central moment to vanish, and $H_2$ obeys the rule to machine precision. Polar molecules break the rule, and the sign is set by orientation, because reversing $\boldsymbol{\lambda}$ is equivalent to the photon parity $\hat{b} \to -\hat{b}$ and therefore changes the sign of every odd central moment exactly. The normalized cumulant $\kappa_{3,\mathrm{c}}/{n_{\mathrm{c}}}^{3/2}$ is $+1.548$ for $HeH^+$ (QED-CCSD-22 versus exact diagonalization, relative error $5.4 \times 10^{-4}$ at $\lambda = 0.02$ a.u., $\omega = 10$ eV) and, at the aug-cc-pVDZ level, $+0.470$ for HF and $-0.133$ for $H_2O$. The coupling vector $\boldsymbol{\lambda}$ points toward the positive end of the charge distribution for $HeH^+$ and HF and away from it for $H_2O$ (geometries in the supplementary material), so the three values agree in sign once the orientation is fixed. This observation does not establish a universal sign relation to the permanent dipole. The third cumulant is thus a signed observable with no second-order counterpart, whose sign interpretation requires molecular polarity information and an optical phase reference.

In an isotropic or head–tail-symmetric ensemble, opposite coupling directions have equal weight and their odd contributions cancel. An individually addressed molecule retains its signal if its orientation remains stable. Chikkaraddy et al. used host–guest alignment in single-molecule strong-coupling experiments,[36] and Gutbrod et al. determined individual transition-dipole orientations in an optical microresonator.[37] These precedents support the orientation setting without demonstrating the signed third-cumulant measurement. Even moments need not vanish, although rotation and geometry relaxation can change the statistics (Sec. S8). For $H_2O$ at $\omega = 10$ eV and $\lambda = 0.02$ a.u., $g_c^{(2)}$ is 12.28 along the $C_2$ axis, 11.49 perpendicular to it within the molecular plane, and 12.53 normal to that plane (Sec. S12, Table S19).

Intracavity-to-emitted relations for cavity-embedded matter have been formulated by Grunwald et al.,[12] providing motivation for a driven input–output treatment. Because the ground state emits nothing, a possible extension is a weak coherent probe of the cavity, with the statistics read from the scattered light. A quantitative prediction requires the driven state, the coupling to the external field, and the detector operator. Plasmonic nanocavities have demonstrated single-molecule strong coupling at room temperature,[36] and plasmonic picocavities reach single-molecule sensitivity in the vibrational domain.[38] As an idealized mode-volume estimate, $\lambda = (4\pi/V_{\mathrm{eff}})^{1/2}$ places the couplings studied here, $\lambda = 0.005$–$0.05$ a.u., at $V_{\mathrm{eff}} \approx 75$–$0.7$ nm$^3$, a nanocavity-to-picocavity scale. The frequency window is set by the spectra of the validation systems, but the statistics depend on frequency through the detuning ratio $\omega/\Delta E$ of the two-level model, suggesting analogous ratios for visible-band chromophores at optical cavity frequencies. The two-level amplitude scales as $\lambda d$, so at fixed detuning ratio $n_{\mathrm{c}} \propto d^2/\Delta E$, and larger transition dipoles and smaller excitation energies could increase $n_c$ within this model.

All results in this study are for the intracavity field of a single lossless cavity mode at fixed nuclear geometry and molecular orientation. Connecting the computed moments to detector-level photon counting requires an input-output treatment, and in the ultrastrong coupling regime, the emitted-light correlations must be constructed from positive-frequency components of the field operator in the dressed eigenbasis, with gauge-invariant open-system master equations.[33,34,39–41] Subtracting the coherent mean does not perform the interacting positive-frequency projection, so the ground-state ratios do not supply the zero-delay value of a detected correlation.

In this work, we identified the displacement-invariant cumulants of the cavity field as convention-independent intracavity photon statistics of a cavity-coupled molecule, computed them from QED-CCSD-22 densities, and validated them against exact Pauli–Fierz diagonalization. Potential next steps include perturbative-triples response densities for strong coupling and frequency-resolved $g^{(2)}(\tau)$ of the cavity output from real-time QED-CC propagation[42] with cavity loss.

## Supplementary Material

See the supplementary material for the working equations for the lab-frame and origin-invariant assemblies, the exact-diagonalization details and Fock-space convergence data, the origin-invariance placement tables, the complete validation, aug-cc-pVDZ, and resonance-scan tables, the third-cumulant values, the invariant quadrature variances and Mandel parameters, and the additional LiH/6-31G benchmark.

## Acknowledgments

This work was supported by the Air Force Office of Scientific Research under AFOSR Award No. FA9550-24-1-0199. We acknowledge the use of artificial intelligence in writing some of the code used in this study and in revising and editing the text.

## Author Declarations

### Conflict of Interest

The author has no conflicts to disclose.

### Author Contributions

Kurt R. Brorsen: Conceptualization (lead), Software (lead), Writing (lead)

**Data Availability**

The data that support the findings of this study are available within the article and its supplementary material. An archival snapshot of the source code, working equations, sweep drivers, and archived outputs is deposited on Zenodo (DOI: 10.5281/zenodo.22033874).

# Supplementary Material for “Intracavity Photon Statistics from Correlated Molecular Electronic Structure”

Kurt R. Brorsen, Department of Chemistry, University of Missouri, Columbia, Missouri 65211, United States (brorsenk@missouri.edu)

### S1. Photon moments

The eight shifted-frame photon moments are defined as

$$M_{\hat{O}} = \langle HF | (1+\hat{\Lambda}) e^{-\hat{T}} \hat{O} e^{\hat{T}} | HF \rangle, \quad \hat{O} \in \{\hat{b}, \hat{b}^{\dagger}, \hat{b}^{2}, \hat{b}^{\dagger 2}, \hat{b}^{\dagger}\hat{b}, \hat{b}^{\dagger}\hat{b}^{2}, \hat{b}^{\dagger 2}\hat{b}, \hat{b}^{\dagger 2}\hat{b}^{2}\}, \tag{S1}$$

where $\hat{T}$ and $\hat{\Lambda}$ are the converged QED-CCSD-22 cluster and de-excitation operators.[1,2] This $\Lambda$-dependent expectation-value structure also underlies one-particle densities and analytic gradients in QED-CC.[3] The expressions were derived symbolically with the SASQ package,[4] with conjugate operator pairs derived independently because CC expectation values are non-Hermitian at finite truncation. Table S1 lists the number of terms per moment, and the $\langle \hat{b}^{\dagger}\hat{b} \rangle$ expression is term-by-term identical to the photon-number block of the validated density-matrix equations. The complete machine-generated expressions are included in the Zenodo repository (see Section S10).

| operator $\hat{O}$ | terms |
|---|---|
| $\hat{b}$ | 8 |
| $\hat{b}^{\dagger}$ | 1 |
| $\hat{b}^{2}$ | 25 |
| $\hat{b}^{\dagger 2}$ | 1 |
| $\hat{b}^{\dagger}\hat{b}$ | 8 |
| $\hat{b}^{\dagger}\hat{b}^{2}$ | 19 |
| $\hat{b}^{\dagger 2}\hat{b}$ | 4 |
| $\hat{b}^{\dagger 2}\hat{b}^{2}$ | 10 |

**TABLE S1**: Number of SASQ-generated contraction terms in each shifted-frame photon-moment expression (76 in total). The asymmetry between conjugate pairs reflects the non-Hermiticity of CC expectation values: $\langle \hat{b}^{\dagger} \rangle$ is the single de-excitation amplitude $\lambda_{g_1}$, whereas $\langle \hat{b} \rangle$ carries $\Lambda$-dressed contributions.

## S2. Frame convention and origin-invariant assembly

The CC equations are solved in the coherent-state frame of the QED-HF reference.[5] The lab and shifted frames are related by the displacement

$$\hat{H}_{shift} = \hat{D}^{\dagger}(z)\hat{H}_{lab}\hat{D}(z),$$
$$\hat{D}(z) = e^{z(\hat{b}^{\dagger} - \hat{b})}, \quad \text{(S2)}$$
$$z = \frac{\langle \lambda \cdot \hat{d} \rangle_{HF}}{\sqrt{2\omega}},$$

so lab-frame operators map onto shifted-frame moments via $\hat{b}_{lab} = \hat{b} + z$. The raw lab-frame quantities

$$\langle \hat{b}^{\dagger}\hat{b} \rangle_{lab} = M_{\hat{b}^{\dagger}\hat{b}} + z(M_{\hat{b}} + M_{\hat{b}^{\dagger}}) + z^2, \quad \text{(S3)}$$

$$\langle \hat{b}^{\dagger}\hat{b}^{\dagger}\hat{b}\hat{b} \rangle_{lab} = M_{\hat{b}^{\dagger 2}\hat{b}^2} + 2z(M_{\hat{b}^{\dagger 2}\hat{b}} + M_{\hat{b}^{\dagger}\hat{b}^2})$$
$$+ z^2(M_{\hat{b}^{\dagger 2}} + M_{\hat{b}^2} + 4M_{\hat{b}^{\dagger}\hat{b}}) + 2z^3(M_{\hat{b}^{\dagger}} + M_{\hat{b}}) + z^4 \quad \text{(S4)}$$

depend on $z$ and are therefore origin- and convention-dependent. The invariant statistics are assembled with the biorthogonal means $\beta = M_{\hat{b}}$ and $\acute{\beta} = M_{\hat{b}^{\dagger}}$, which are independent numbers at finite truncation:

$$n_c = M_{\hat{b}^{\dagger}\hat{b}} - \acute{\beta}\beta, \quad \text{(S5)}$$

$$G_c^{(2)} = M_{\hat{b}^{\dagger 2}\hat{b}^2} - 2\beta M_{\hat{b}^{\dagger 2}\hat{b}} - 2\acute{\beta} M_{\hat{b}^{\dagger}\hat{b}^2}$$
$$+ \beta^2 M_{\hat{b}^{\dagger 2}} + \acute{\beta}^2 M_{\hat{b}^2} + 4\acute{\beta}\beta M_{\hat{b}^{\dagger}\hat{b}} - 3\acute{\beta}^2\beta^2, \quad \text{(S6)}$$

$$\kappa_{3,c} = M_{\hat{b}^{\dagger}\hat{b}^2} - 2\beta M_{\hat{b}^{\dagger}\hat{b}} - \acute{\beta} M_{\hat{b}^2} + 2\acute{\beta}\beta^2, \quad \text{(S7)}$$

with $g_c^{(2)} = G_c^{(2)} / n_c{}^2$ and $\acute{\kappa}_3 = \kappa_{3,c} / n_c{}^{3/2}$. The coherent shift $z$ is absent from Eqs. (S5)–(S7), so the displacement invariance of the main text is directly obtained from the working equations rather than achieved by numerical cancellation. All three expressions vanish when every moment factorizes over a coherent state, which is the mean-field null result. The normalized ratios are undefined at this $n_c = 0$ limit.

Finite biorthogonal CC does not guarantee a positive semidefinite photon density operator or the usual conjugation relations between moments. Mean subtraction preserves displacement invariance without restoring positivity. All 48 original validation points in Tables S4–S6, 68 molecular/basis-scan points in Tables S7–S11, four new LiH/6-31G points in Table S18, and the three $H_2O$ orientations in Table S19 have positive $n_c$ and $g_c^{(2)}$, and hence positive $G_c^{(2)} = n_c^2 g_c^{(2)}$. These numerical checks do not prove that every constraint on a physical density operator is satisfied. A negative incoherent photon number or fourth-order central moment would indicate failure of the approximation for that property.

## S3. Exact Pauli–Fierz reference

The reference calculations are obtained by diagonalizing the exact-DSE Pauli–Fierz Hamiltonian in the product basis of the full configuration-interaction electronic space and a photon Fock space truncated at $n_{max}$, in the lab frame, with the identical bare-electronic-position dipole convention as the CC code. With $|\psi\rangle$ the (real) ground vector and $\delta\hat{b}=\hat{b}-\langle\psi|\hat{b}|\psi\rangle$, the invariant statistics are evaluated as vector norms,

$$\begin{aligned} n_c &= \|\delta\hat{b}|\psi\rangle\|^2, \\ G_c^{(2)} &= \|\delta\hat{b}^2|\psi\rangle\|^2, \\ \kappa_{3,c} &= \langle\delta\hat{b}\psi|\delta\hat{b}^2\psi\rangle. \end{aligned} \tag{S8}$$

All exact-diagonalization values in the main text are computed using $n_{max}=20$. Table S2 presents the $n_{max}$ convergence at $\lambda=0.05$ a.u. and $\omega=10$ eV in the lab frame with the electronic-position dipole convention. The $n_{max}=12\rightarrow20$ change is below $2\times10^{-9}$ relative. Fock truncation at $n_{max}=2$ misestimates $g_c^{(2)}$ by $+11\%$ for $H_2$ and by a factor of 12 for LiH while $n\approx6$–8 is required for convergence in these examples. The required cutoff depends on coupling, frequency, coherent displacement, and the requested accuracy, so the chosen statistic must be converged in its chosen frame. A coherent displacement can reduce the required photon space. Disconnected powers of the QED-CCSD-22 exponential populate photon sectors above two, so its photon excitation rank is distinct from a hard Fock cutoff.

| system | $n_{max}$ | $g_c^{(2)}$ | rel. error |
|---|---|---|---|
| $H_2$/6-31G | 2 | 9.20663705 | $+1.096\times10^{-1}$ |
| | 4 | 8.29716336 | $-4.104\times10^{-6}$ |
| | 6 | 8.29719740 | $-9.309\times10^{-10}$ |
| | 8 | 8.29719741 | $-3.425\times10^{-14}$ |
| | 12 | 8.29719741 | $+1.477\times10^{-14}$ |
| | 16 | 8.29719741 | $-7.279\times10^{-15}$ |
| | 20 | 8.29719741 | $+0.000\times10^{0}$ |
| $HeH^+$/STO-3G | 2 | 17.14151355 | $-6.397\times10^{-3}$ |
| | 4 | 17.25187940 | $-1.730\times10^{-7}$ |
| | 6 | 17.25188238 | $-1.460\times10^{-12}$ |
| | 8 | 17.25188238 | $-5.890\times10^{-14}$ |
| | 12 | 17.25188238 | $-5.498\times10^{-14}$ |
| | 16 | 17.25188238 | $-2.306\times10^{-13}$ |
| | 20 | 17.25188238 | $+0.000\times10^{0}$ |
| LiH/STO-3G | 2 | 208.81632427 | $+1.142\times10^{1}$ |
| | 4 | 17.20055204 | $+2.306\times10^{-2}$ |
| | 6 | 16.81308329 | $+1.107\times10^{-5}$ |
| | 8 | 16.81289715 | $+2.692\times10^{-9}$ |
| | 12 | 16.81289710 | $-1.606\times10^{-14}$ |
| | 16 | 16.81289710 | $-2.219\times10^{-14}$ |
| | 20 | 16.81289710 | $+0.000\times10^{0}$ |

**TABLE S2**: Convergence of the exact-diagonalization $g_c^{(2)}$ with the photon Fock-space truncation $n_{max}$ ($\lambda=0.05$ a.u. along the bond, $\omega=10$ eV; lab frame, electronic-position dipole convention). Relative errors are referenced against $n_{max}=20$.

## S4. Origin-invariance placement data

Table S3 shows that upon rigid translation of the system for $H_2$/STO-3G and $HeH^+$/STO-3G, the raw $g^{(2)}(0)$ varies from 1.00 to 15.2 across placements, while $g_c^{(2)}$ is constant to a relative spread of $2.6\times10^{-12}$ for $H_2$ and $5.2\times10^{-10}$ for $HeH^+$. Note that the STO-3G $g_c^{(2)}$ of $H_2$ (15.17 at $\lambda=0.05$ a.u., $\omega=10$ eV) differs from the 6-31G value quoted elsewhere. The invariance demonstration is basis-independent.

| Geometry | $z$ | raw $g^{(2)}(0)$ | $g_c^{(2)}$ |
|---|---|---|---|
| H 0 0 0; H 0.74 0 0 | $+0.0816$ | 1.469818 | 15.1743059010 |
| H −0.37 0 0; H 0.37 0 0 | $-0.0000$ | 15.174306 | 15.1743059010 |
| H 5.0 0 0; H 5.74 0 0 | $+1.1837$ | 1.002286 | 15.1743059010 |
| | | | |
| He −5 0 0; H −4.225 0 0 | $-1.0820$ | 1.001394 | 17.2192357984 |
| He −2.5 0 0; H −1.725 0 0 | $-0.5309$ | 1.005767 | 17.2192358038 |
| He 0 0 0; H 0.775 0 0 | $+0.0201$ | 4.505933 | 17.2192358074 |
| He 2.5 0 0; H 3.275 0 0 | $+0.5712$ | 1.005048 | 17.2192358047 |
| He 5 0 0; H 5.775 0 0 | $+1.1222$ | 1.001304 | 17.2192358050 |

**TABLE S3**: Demonstration of the origin-invariance of QED-CCSD-22 ($\lambda=0.05$ a.u. along the bond, $\omega=10$ eV; $H_2$/STO-3G upper block, $HeH^+$/STO-3G lower block). Rigid translations change the coherent shift $z$ and the raw $g^{(2)}(0)$, while the invariant $g_c^{(2)}$ is placement-independent. Coordinates in Å.

## S5. Validation sweep

Tables S4–S6 show the complete $(\lambda,\omega)$ validation sweep against exact diagonalization. Geometries: $H_2$ ($R=0.74$ Å, 6-31G), $HeH^+$ ($R=0.775$ Å, STO-3G), LiH ($R=1.595$ Å, STO-3G), with $\lambda$ along the bond. For the two-electron systems the electronic treatment is exact and the $\Delta E$ column isolates the truncation of the photon excitation rank. LiH carries an additional $\lambda$-independent electronic error floor. The $\lambda=0.1$ a.u. endpoint tests the limits of the approximation, while the aug-cc-pVDZ frequency scan in Sec. S6 uses 0.02 a.u. Additional LiH/6-31G validation and photon-cutoff checks are given in Sec. S11.

| $\lambda$ | $\omega$ (eV) | $\Delta E$ | $n_c$ (ex.) | $g_c^{(2)}$ (ex.) | $g_c^{(2)}$ (CC) | rel. |
|---|---|---|---|---|---|---|
| 0.005 | 5.0 | $-4.4\times10^{-16}$ | $6.9995\times10^{-6}$ | 18.59480 | 18.59413 | $3.6\times10^{-5}$ |
| 0.005 | 10.0 | $+5.1\times10^{-15}$ | $9.0328\times10^{-6}$ | 8.28374 | 8.28339 | $4.3\times10^{-5}$ |
| 0.005 | 15.0 | $+8.2\times10^{-15}$ | $9.4654\times10^{-6}$ | 5.90129 | 5.90104 | $4.2\times10^{-5}$ |
| 0.005 | 20.0 | $-8.0\times10^{-15}$ | $9.3153\times10^{-6}$ | 4.90540 | 4.90521 | $3.9\times10^{-5}$ |
| 0.010 | 5.0 | $-1.3\times10^{-15}$ | $2.7996\times10^{-5}$ | 18.59161 | 18.58895 | $1.4\times10^{-4}$ |
| 0.010 | 10.0 | $-2.2\times10^{-14}$ | $3.6126\times10^{-5}$ | 8.28416 | 8.28274 | $1.7\times10^{-4}$ |
| 0.010 | 15.0 | $-5.3\times10^{-14}$ | $3.7857\times10^{-5}$ | 5.90176 | 5.90077 | $1.7\times10^{-4}$ |
| 0.010 | 20.0 | $-3.8\times10^{-14}$ | $3.7257\times10^{-5}$ | 4.90579 | 4.90502 | $1.6\times10^{-4}$ |
| 0.020 | 5.0 | $-9.5\times10^{-13}$ | $1.1195\times10^{-4}$ | 18.57888 | 18.56822 | $5.7\times10^{-4}$ |
| 0.020 | 10.0 | $-2.4\times10^{-12}$ | $1.4442\times10^{-4}$ | 8.28581 | 8.28015 | $6.8\times10^{-4}$ |
| 0.020 | 15.0 | $-2.8\times10^{-12}$ | $1.5135\times10^{-4}$ | 5.90363 | 5.89968 | $6.7\times10^{-4}$ |
| 0.020 | 20.0 | $-2.8\times10^{-12}$ | $1.4896\times10^{-4}$ | 4.90732 | 4.90425 | $6.3\times10^{-4}$ |
| 0.050 | 5.0 | $-4.9\times10^{-10}$ | $6.9823\times10^{-4}$ | 18.49037 | 18.42428 | $3.6\times10^{-3}$ |
| 0.050 | 10.0 | $-8.3\times10^{-10}$ | $8.9915\times10^{-4}$ | 8.29720 | 8.26194 | $4.2\times10^{-3}$ |
| 0.050 | 15.0 | $-9.4\times10^{-10}$ | $9.4253\times10^{-4}$ | 5.91668 | 5.89201 | $4.2\times10^{-3}$ |
| 0.050 | 20.0 | $-9.3\times10^{-10}$ | $9.2808\times10^{-4}$ | 4.91803 | 4.89885 | $3.9\times10^{-3}$ |
| 0.100 | 5.0 | $-2.7\times10^{-8}$ | $2.7720\times10^{-3}$ | 18.18319 | 17.92607 | $1.4\times10^{-2}$ |
| 0.100 | 10.0 | $-4.7\times10^{-8}$ | $3.5478\times10^{-3}$ | 8.33569 | 8.19607 | $1.7\times10^{-2}$ |
| 0.100 | 15.0 | $-5.5\times10^{-8}$ | $3.7223\times10^{-3}$ | 5.96214 | 5.86405 | $1.6\times10^{-2}$ |
| 0.100 | 20.0 | $-5.5\times10^{-8}$ | $3.6712\times10^{-3}$ | 4.95564 | 4.87927 | $1.5\times10^{-2}$ |

**TABLE S4**: Validation sweep for $H_2$/6-31G comparing QED-CCSD-22 vs exact Pauli–Fierz diagonalization ($n_{max}=20$; $\lambda$ along the bond). $\Delta E=E_{CC}-E_{exact}$ in Hartree. rel. $=|g_{c,CC}^{(2)}-g_{c,ex}^{(2)}|/g_{c,ex}^{(2)}$.

| $\lambda$ | $\omega$ (eV) | $\Delta E$ | $n_c$ (ex.) | $g_c^{(2)}$ (ex.) | $g_c^{(2)}$ (CC) | rel. |
|---|---|---|---|---|---|---|
| 0.005 | 5.0 | $-7.0\times10^{-13}$ | $1.1303\times10^{-6}$ | 46.93399 | 46.93333 | $1.4\times10^{-5}$ |
| 0.005 | 10.0 | $-7.0\times10^{-13}$ | $1.7053\times10^{-6}$ | 17.28467 | 17.28434 | $1.9\times10^{-5}$ |
| 0.005 | 15.0 | $-7.0\times10^{-13}$ | $1.9981\times10^{-6}$ | 10.78028 | 10.78005 | $2.0\times10^{-5}$ |
| 0.005 | 20.0 | $-7.0\times10^{-13}$ | $2.1384\times10^{-6}$ | 8.14302 | 8.14285 | $2.1\times10^{-5}$ |
| 0.010 | 5.0 | $-3.3\times10^{-13}$ | $4.5216\times10^{-6}$ | 46.92496 | 46.92231 | $5.6\times10^{-5}$ |
| 0.010 | 10.0 | $-3.3\times10^{-13}$ | $6.8213\times10^{-6}$ | 17.28367 | 17.28236 | $7.6\times10^{-5}$ |
| 0.010 | 15.0 | $-3.3\times10^{-13}$ | $7.9924\times10^{-6}$ | 10.78005 | 10.77917 | $8.2\times10^{-5}$ |
| 0.010 | 20.0 | $-3.3\times10^{-13}$ | $8.5537\times10^{-6}$ | 8.14297 | 8.14229 | $8.3\times10^{-5}$ |
| 0.020 | 5.0 | $-3.5\times10^{-13}$ | $1.8091\times10^{-5}$ | 46.88889 | 46.87832 | $2.3\times10^{-4}$ |
| 0.020 | 10.0 | $-2.6\times10^{-13}$ | $2.7286\times10^{-5}$ | 17.27968 | 17.27444 | $3.0\times10^{-4}$ |
| 0.020 | 15.0 | $-2.3\times10^{-13}$ | $3.1968\times10^{-5}$ | 10.77916 | 10.77563 | $3.3\times10^{-4}$ |
| 0.020 | 20.0 | $-2.3\times10^{-13}$ | $3.4213\times10^{-5}$ | 8.14277 | 8.14007 | $3.3\times10^{-4}$ |
| 0.050 | 5.0 | $+3.4\times10^{-11}$ | $1.1326\times10^{-4}$ | 46.63811 | 46.57255 | $1.4\times10^{-3}$ |
| 0.050 | 10.0 | $+5.6\times10^{-11}$ | $1.7056\times10^{-4}$ | 17.25188 | 17.21924 | $1.9\times10^{-3}$ |
| 0.050 | 15.0 | $+6.3\times10^{-11}$ | $1.9975\times10^{-4}$ | 10.77295 | 10.75091 | $2.0\times10^{-3}$ |
| 0.050 | 20.0 | $+6.2\times10^{-11}$ | $2.1375\times10^{-4}$ | 8.14140 | 8.12456 | $2.1\times10^{-3}$ |
| 0.100 | 5.0 | $+2.3\times10^{-9}$ | $4.5572\times10^{-4}$ | 45.76669 | 45.51097 | $5.6\times10^{-3}$ |
| 0.100 | 10.0 | $+3.7\times10^{-9}$ | $6.8251\times10^{-4}$ | 17.15433 | 17.02514 | $7.5\times10^{-3}$ |
| 0.100 | 15.0 | $+4.1\times10^{-9}$ | $7.9831\times10^{-4}$ | 10.75119 | 10.66360 | $8.1\times10^{-3}$ |
| 0.100 | 20.0 | $+4.1\times10^{-9}$ | $8.5390\times10^{-4}$ | 8.13669 | 8.06962 | $8.2\times10^{-3}$ |

**TABLE S5**: Validation sweep for $HeH^+$/STO-3G comparing QED-CCSD-22 vs exact Pauli–Fierz diagonalization ($n_{max}=20$; $\lambda$ along the bond). $\Delta E=E_{CC}-E_{exact}$ in Hartree. rel. $=|g^{(2)}_{c,CC}-g^{(2)}_{c,ex}|/g^{(2)}_{c,ex}$.

| $\lambda$ | $\omega$ (eV) | $\Delta E$ | $n_c$ (ex.) | $g_c^{(2)}$ (ex.) | $g_c^{(2)}$ (CC) | rel. |
|---|---|---|---|---|---|---|
| 0.005 | 10.0 | $+1.0\times10^{-5}$ | $1.3135\times10^{-5}$ | 17.20855 | 17.20224 | $3.7\times10^{-4}$ |
| 0.010 | 10.0 | $+1.1\times10^{-5}$ | $5.2601\times10^{-5}$ | 17.19626 | 17.15090 | $2.6\times10^{-3}$ |
| 0.020 | 10.0 | $+1.1\times10^{-5}$ | $2.1139\times10^{-4}$ | 17.14727 | 16.94823 | $1.2\times10^{-2}$ |
| 0.050 | 5.0 | $+1.1\times10^{-5}$ | $1.5070\times10^{-3}$ | 25.96874 | 23.29722 | $1.0\times10^{-1}$ |
| 0.050 | 10.0 | $+1.1\times10^{-5}$ | $1.3634\times10^{-3}$ | 16.81290 | 15.64265 | $7.0\times10^{-2}$ |
| 0.050 | 15.0 | $+1.1\times10^{-5}$ | $1.1872\times10^{-3}$ | 13.25988 | 12.57098 | $5.2\times10^{-2}$ |
| 0.050 | 20.0 | $+1.1\times10^{-5}$ | $1.0451\times10^{-3}$ | 11.34004 | 10.87172 | $4.1\times10^{-2}$ |
| 0.100 | 10.0 | $+2.7\times10^{-5}$ | $5.9901\times10^{-3}$ | 15.75840 | 12.28311 | $2.2\times10^{-1}$ |

**TABLE S6**: Validation sweep for LiH/STO-3G comparing QED-CCSD-22 vs exact Pauli–Fierz diagonalization ($n_{max}=20$; $\lambda$ along the bond). $\Delta E=E_{CC}-E_{exact}$ in Hartree. rel. $=|g^{(2)}_{c,CC}-g^{(2)}_{c,ex}|/g^{(2)}_{c,ex}$.

## S6. aug-cc-pVDZ sweep

Tables S7–S11 show the complete aug-cc-pVDZ scan (QED-CCSD-22 values only). Geometries are from Ref. 6: $H_2$ (H $\pm 0.3707$ on $x$), HF (F at the origin, H at $z=0.9168$), $H_2O$ (O at $z=0.1173$, H at $y=\pm 0.7572$, $z=-0.4692$) (coordinates in Å). The coordinates of $HeH^+$ and LiH are the same as the validation geometries. The coupling vector lies along the molecular axis for the diatomics and along the $C_{2v}$ axis for $H_2O$. For each system, we performed a $\omega$ scan at $\lambda=0.02$ a.u., a $\lambda$ scan at $\omega=10$ eV, and, for $H_2$, HF, and $H_2O$, calculations using the cc-pVDZ and aug-cc-pVTZ basis sets. Table S13 presents the second-diffuse-shell check (aug-cc-pVDZ → d-aug-cc-pVDZ) for the same three systems at $\lambda=0.02$ a.u. For the two-electron systems, exact diagonalization (Sec. S3) is possible at aug-cc-pVDZ. Table S12 validates the archived $H_2$ and $HeH^+$ aug-cc-pVDZ values against exact diagonalization at every frequency of the $\omega$ scan, with a Fock-space convergence check at $\omega=10$ eV.

| basis | $N_{bf}$ | $\lambda$ | $\omega$ (eV) | $E_{CC}$ | $n_c$ | $g_c^{(2)}$ |
|---|---|---|---|---|---|---|
| aug-cc-pvdz | 18 | 0.005 | 10.0 | -1.16461018 | $9.8344\times10^{-6}$ | 10.05302 |
| aug-cc-pvdz | 18 | 0.010 | 10.0 | -1.16457061 | $3.9332\times10^{-5}$ | 10.04769 |
| aug-cc-pvdz | 18 | 0.020 | 5.0 | -1.16436016 | $1.2553\times10^{-4}$ | 19.62138 |
| aug-cc-pvdz | 18 | 0.020 | 7.5 | -1.16438923 | $1.4691\times10^{-4}$ | 12.84487 |
| aug-cc-pvdz | 18 | 0.020 | 10.0 | -1.16441237 | $1.5723\times10^{-4}$ | 10.02645 |
| aug-cc-pvdz | 18 | 0.020 | 12.5 | -1.16443126 | $1.6134\times10^{-4}$ | 8.50664 |
| aug-cc-pvdz | 18 | 0.020 | 15.0 | -1.16444698 | $1.6185\times10^{-4}$ | 7.56063 |
| aug-cc-pvdz | 18 | 0.020 | 17.5 | -1.16446027 | $1.6025\times10^{-4}$ | 6.91600 |
| aug-cc-pvdz | 18 | 0.020 | 20.0 | -1.16447166 | $1.5742\times10^{-4}$ | 6.44860 |
| aug-cc-pvdz | 18 | 0.050 | 10.0 | -1.16330637 | $9.7862\times10^{-4}$ | 9.88261 |
| aug-cc-pvtz | 46 | 0.020 | 5.0 | -1.17237489 | $1.2247\times10^{-4}$ | 19.99875 |
| aug-cc-pvtz | 46 | 0.020 | 10.0 | -1.17242626 | $1.5375\times10^{-4}$ | 10.18135 |
| aug-cc-pvtz | 46 | 0.020 | 20.0 | -1.17248472 | $1.5432\times10^{-4}$ | 6.48272 |
| cc-pvdz | 10 | 0.020 | 5.0 | -1.16315293 | $1.2043\times10^{-4}$ | 17.08313 |
| cc-pvdz | 10 | 0.020 | 10.0 | -1.16320402 | $1.5179\times10^{-4}$ | 7.89484 |
| cc-pvdz | 10 | 0.020 | 20.0 | -1.16326227 | $1.5281\times10^{-4}$ | 4.79545 |

**TABLE S7:** aug-cc-pVDZ sweep for $H_2$ (QED-CCSD-22 values only; Weber geometries, $\lambda$ along the molecular/$C_{2v}$ axis). $E_{CC}$ in Hartree.

| basis | $N_{bf}$ | $\lambda$ | $\omega$ (eV) | $E_{CC}$ | $n_c$ | $g_c^{(2)}$ |
|---|---|---|---|---|---|---|
| aug-cc-pvdz | 32 | 0.005 | 10.0 | -100.26147801 | $8.2757\times10^{-6}$ | 14.35700 |
| aug-cc-pvdz | 32 | 0.010 | 10.0 | -100.26139338 | $3.3100\times10^{-5}$ | 14.34997 |
| aug-cc-pvdz | 32 | 0.020 | 5.0 | -100.26098975 | $9.5544\times10^{-5}$ | 33.07970 |
| aug-cc-pvdz | 32 | 0.020 | 7.5 | -100.26102498 | $1.1809\times10^{-4}$ | 19.70147 |
| aug-cc-pvdz | 32 | 0.020 | 10.0 | -100.26105494 | $1.3235\times10^{-4}$ | 14.32196 |
| aug-cc-pvdz | 32 | 0.020 | 12.5 | -100.26108080 | $1.4130\times10^{-4}$ | 11.49942 |
| aug-cc-pvdz | 32 | 0.020 | 15.0 | -100.26110341 | $1.4673\times10^{-4}$ | 9.78177 |
| aug-cc-pvdz | 32 | 0.020 | 17.5 | -100.26112339 | $1.4979\times10^{-4}$ | 8.63345 |
| aug-cc-pvdz | 32 | 0.020 | 20.0 | -100.26114120 | $1.5120\times10^{-4}$ | 7.81451 |
| aug-cc-pvdz | 32 | 0.050 | 10.0 | -100.25868897 | $8.2514\times10^{-4}$ | 14.13110 |
| aug-cc-pvtz | 69 | 0.020 | 5.0 | -100.35546999 | $9.6984\times10^{-5}$ | 32.92547 |
| aug-cc-pvtz | 69 | 0.020 | 10.0 | -100.35553578 | $1.3438\times10^{-4}$ | 14.23897 |
| aug-cc-pvtz | 69 | 0.020 | 20.0 | -100.35562259 | $1.5345\times10^{-4}$ | 7.76287 |
| cc-pvdz | 19 | 0.020 | 5.0 | -100.22763971 | $5.9634\times10^{-5}$ | 37.63940 |
| cc-pvdz | 19 | 0.020 | 10.0 | -100.22768531 | $8.3309\times10^{-5}$ | 15.74956 |
| cc-pvdz | 19 | 0.020 | 20.0 | -100.22774820 | $9.7190\times10^{-5}$ | 8.22108 |

**TABLE S8:** aug-cc-pVDZ sweep for HF (QED-CCSD-22 values only; Weber geometries, $\lambda$ along the molecular/$C_{2v}$ axis). $E_{CC}$ in Hartree.

| basis | $N_{bf}$ | $\lambda$ | $\omega$ (eV) | $E_{CC}$ | $n_c$ | $g_c^{(2)}$ |
|---|---|---|---|---|---|---|
| aug-cc-pvdz | 41 | 0.005 | 10.0 | -76.27075340 | $1.2450\times10^{-5}$ | 12.31402 |
| aug-cc-pvdz | 41 | 0.010 | 10.0 | -76.27065250 | $4.9794\times10^{-5}$ | 12.30624 |
| aug-cc-pvdz | 41 | 0.020 | 5.0 | -76.27016324 | $1.5193\times10^{-4}$ | 26.33651 |
| aug-cc-pvdz | 41 | 0.020 | 7.5 | -76.27020997 | $1.8186\times10^{-4}$ | 16.41771 |
| aug-cc-pvdz | 41 | 0.020 | 10.0 | -76.27024902 | $1.9907\times10^{-4}$ | 12.27529 |
| aug-cc-pvdz | 41 | 0.020 | 12.5 | -76.27028227 | $2.0872\times10^{-4}$ | 10.04309 |
| aug-cc-pvdz | 41 | 0.020 | 15.0 | -76.27031104 | $2.1365\times10^{-4}$ | 8.65887 |
| aug-cc-pvdz | 41 | 0.020 | 17.5 | -76.27033622 | $2.1556\times10^{-4}$ | 7.72094 |
| aug-cc-pvdz | 41 | 0.020 | 20.0 | -76.27035849 | $2.1548\times10^{-4}$ | 7.04553 |
| aug-cc-pvdz | 41 | 0.050 | 10.0 | -76.26742999 | $1.2394\times10^{-3}$ | 12.06618 |
| aug-cc-pvtz | 92 | 0.020 | 5.0 | -76.34815082 | $1.5960\times10^{-4}$ | 25.94644 |
| aug-cc-pvtz | 92 | 0.020 | 10.0 | -76.34823899 | $2.0884\times10^{-4}$ | 12.18593 |
| aug-cc-pvtz | 92 | 0.020 | 20.0 | -76.34835053 | $2.2487\times10^{-4}$ | 7.08393 |
| cc-pvdz | 24 | 0.020 | 5.0 | -76.23947672 | $7.9179\times10^{-5}$ | 31.72466 |
| cc-pvdz | 24 | 0.020 | 10.0 | -76.23953147 | $1.0587\times10^{-4}$ | 14.00800 |
| cc-pvdz | 24 | 0.020 | 20.0 | -76.23960653 | $1.1956\times10^{-4}$ | 7.66406 |

**TABLE S9:** aug-cc-pVDZ sweep for $H_2O$ (QED-CCSD-22 values only; Weber geometries, $\lambda$ along the molecular/$C_{2v}$ axis). $E_{CC}$ in Hartree.

| basis | $N_{bf}$ | $\lambda$ | $\omega$ (eV) | $E_{CC}$ | $n_c$ | $g_c^{(2)}$ |
|---|---|---|---|---|---|---|
| aug-cc-pvdz | 18 | 0.005 | 10.0 | -2.96171426 | $1.9507\times10^{-6}$ | 17.97822 |
| aug-cc-pvdz | 18 | 0.010 | 10.0 | -2.96169037 | $7.8028\times10^{-6}$ | 17.97551 |
| aug-cc-pvdz | 18 | 0.020 | 5.0 | -2.96157647 | $2.0844\times10^{-5}$ | 46.69977 |
| aug-cc-pvdz | 18 | 0.020 | 7.5 | -2.96158631 | $2.6906\times10^{-5}$ | 25.91874 |
| aug-cc-pvdz | 18 | 0.020 | 10.0 | -2.96159484 | $3.1208\times10^{-5}$ | 17.96468 |
| aug-cc-pvdz | 18 | 0.020 | 12.5 | -2.96160233 | $3.4255\times10^{-5}$ | 13.95804 |
| aug-cc-pvdz | 18 | 0.020 | 15.0 | -2.96160895 | $3.6391\times10^{-5}$ | 11.60011 |
| aug-cc-pvdz | 18 | 0.020 | 17.5 | -2.96161484 | $3.7857\times10^{-5}$ | 10.06650 |
| aug-cc-pvdz | 18 | 0.020 | 20.0 | -2.96162013 | $3.8825\times10^{-5}$ | 8.99727 |
| aug-cc-pvdz | 18 | 0.050 | 10.0 | -2.96092652 | $1.9493\times10^{-4}$ | 17.88961 |

**TABLE S10:** aug-cc-pVDZ sweep for $HeH^+$ (QED-CCSD-22 values only; Weber geometries, $\lambda$ along the molecular/$C_{2v}$ axis). $E_{CC}$ in Hartree.

| basis | $N_{bf}$ | $\lambda$ | $\omega$ (eV) | $E_{CC}$ | $n_c$ | $g_c^{(2)}$ |
|---|---|---|---|---|---|---|
| aug-cc-pvdz | 32 | 0.005 | 10.0 | -8.02119866 | $3.4814\times10^{-5}$ | 8.57087 |
| aug-cc-pvdz | 32 | 0.010 | 10.0 | -8.02113758 | $1.3919\times10^{-4}$ | 8.53811 |
| aug-cc-pvdz | 32 | 0.020 | 5.0 | -8.02079901 | $6.0495\times10^{-4}$ | 12.94891 |
| aug-cc-pvdz | 32 | 0.020 | 7.5 | -8.02085410 | $5.8941\times10^{-4}$ | 9.92994 |
| aug-cc-pvdz | 32 | 0.020 | 10.0 | -8.02089338 | $5.5560\times10^{-4}$ | 8.41091 |
| aug-cc-pvdz | 32 | 0.020 | 12.5 | -8.02092307 | $5.1925\times10^{-4}$ | 7.49486 |
| aug-cc-pvdz | 32 | 0.020 | 15.0 | -8.02094643 | $4.8481\times10^{-4}$ | 6.88227 |
| aug-cc-pvdz | 32 | 0.020 | 17.5 | -8.02096535 | $4.5343\times10^{-4}$ | 6.44392 |
| aug-cc-pvdz | 32 | 0.020 | 20.0 | -8.02098103 | $4.2525\times10^{-4}$ | 6.11482 |
| aug-cc-pvdz | 32 | 0.050 | 10.0 | -8.01918590 | $3.4202\times10^{-3}$ | 7.66215 |

**TABLE S11:** aug-cc-pVDZ sweep for LiH (QED-CCSD-22 values only; Weber geometries, $\lambda$ along the molecular/$C_{2v}$ axis). $E_{CC}$ in Hartree.

| system | $\omega$ (eV) | $g_c^{(2)}$ (exact) | $g_c^{(2)}$ (CC) | rel. dev. |
|---|---|---|---|---|
| $H_2$ | 5.0 | 19.66203 | 19.62138 | $2.07\times10^{-3}$ |
| $H_2$ | 7.5 | 12.87454 | 12.84487 | $2.30\times10^{-3}$ |
| $H_2$ | 10.0 | 10.04971 | 10.02645 | $2.31\times10^{-3}$ |
| $H_2$ | 12.5 | 8.52570 | 8.50664 | $2.24\times10^{-3}$ |
| $H_2$ | 15.0 | 7.57675 | 7.56063 | $2.13\times10^{-3}$ |
| $H_2$ | 17.5 | 6.92994 | 6.91600 | $2.01\times10^{-3}$ |
| $H_2$ | 20.0 | 6.46087 | 6.44860 | $1.90\times10^{-3}$ |
| $HeH^+$ | 5.0 | 46.71254 | 46.69977 | $2.73\times10^{-4}$ |
| $HeH^+$ | 7.5 | 25.92832 | 25.91874 | $3.70\times10^{-4}$ |
| $HeH^+$ | 10.0 | 17.97248 | 17.96468 | $4.34\times10^{-4}$ |
| $HeH^+$ | 12.5 | 13.96466 | 13.95804 | $4.74\times10^{-4}$ |
| $HeH^+$ | 15.0 | 11.60586 | 11.60011 | $4.96\times10^{-4}$ |
| $HeH^+$ | 17.5 | 10.07159 | 10.06650 | $5.06\times10^{-4}$ |
| $HeH^+$ | 20.0 | 9.00184 | 8.99727 | $5.08\times10^{-4}$ |

**TABLE S12:** Validation of the aug-cc-pVDZ values by exact diagonalization: $g_c^{(2)}$ from exact Pauli–Fierz diagonalization at aug-cc-pVDZ ($n_{max}=20$) against the QED-CCSD-22 aug-cc-pVDZ values ($\lambda=0.02$ a.u. along the molecular axis). Exact diagonalization is feasible at this basis only for the two-electron systems. Raising $n_{max}$ from 16 to 20 changes the exact $g_c^{(2)}$ at $\omega=10$ eV by $2.9\times10^{-14}$ ($H_2$) and $7.5\times10^{-13}$ ($HeH^+$).

| system | $\omega$ (eV) | $g_c^{(2)}$ (aVDZ) | $g_c^{(2)}$ (d-aug) | shift |
|---|---|---|---|---|
| $H_2$ | 5.0 | 19.62138 | 19.79799 | $+0.90\%$ |
| $H_2$ | 10.0 | 10.02645 | 10.13756 | $+1.11\%$ |
| $H_2$ | 20.0 | 6.44860 | 6.51185 | $+0.98\%$ |
| HF | 5.0 | 33.07970 | 32.40224 | $-2.05\%$ |
| HF | 10.0 | 14.32196 | 14.17968 | $-0.99\%$ |
| HF | 20.0 | 7.81451 | 7.84001 | $+0.33\%$ |
| $H_2O$ | 5.0 | 26.33651 | 25.00584 | $-5.05\%$ |
| $H_2O$ | 10.0 | 12.27529 | 11.96867 | $-2.50\%$ |
| $H_2O$ | 20.0 | 7.04553 | 7.04080 | $-0.07\%$ |

**TABLE S13:** Second-diffuse-shell check **of** $g_c^{(2)}$ at aug-cc-pVDZ versus d-aug-cc-pVDZ ($\lambda = 0.02$ a.u. along the molecular/$C_{2v}$ axis; aVDZ values from the aug-cc-pVDZ sweep). The largest shift is 5.05%.

## S7. Resonance scan

Table S14 shows the FCI excitation energies and transition-dipole magnitudes that determine the bright $\sigma \rightarrow \sigma^*$ state at 15.309 eV in the one-electron space of the scan. Table S15 shows all 61 points of the fine cavity-frequency scan of the main text. The third cumulant was evaluated at every grid point as a consistency check on the parity selection rule: $|\kappa_{3,c}| \leq 1.8 \times 10^{-16}$ (CC) and $\leq 3.3 \times 10^{-18}$ (exact) across the scan.

| root | $\Delta E$ (eV) | $\lvert\mu_x\rvert$ (a.u.) |
|---|---|---|
| 1 | 10.7419 | 0.0000 |
| 2 | 15.3090 | 1.2978 |
| 3 | 23.3985 | 0.0000 |
| 4 | 28.4986 | 0.0000 |
| 5 | 30.2332 | 0.0000 |

**TABLE S14:** Electronic FCI spectrum of $H_2$/6-31G ($R = 0.74$ Å) in the one-electron space of the resonance scan: excitation energies and transition-dipole magnitudes along the bond ($x$) axis. Root 2 is the bright $\sigma \rightarrow \sigma^*$ state at 15.3090 eV.

| $\omega$ | $n_c \times 10^4$ | $g_c^{(2)}$ (ex.) | $g_c^{(2)}$ (CC) | $\omega$ | $n_c \times 10^4$ | $g_c^{(2)}$ (ex.) | $g_c^{(2)}$ (CC) |
|---|---|---|---|---|---|---|---|
| 5.00 | 6.98226 | 18.49037 | 18.42428 | 12.75 | 9.33809 | 6.69840 | 6.66995 |
| 5.25 | 7.15267 | 17.32448 | 17.26130 | 13.00 | 9.35484 | 6.59462 | 6.56665 |
| 5.50 | 7.31311 | 16.29992 | 16.23939 | 13.25 | 9.36962 | 6.49574 | 6.46824 |
| 5.75 | 7.46412 | 15.39393 | 15.33583 | 13.50 | 9.38252 | 6.40144 | 6.37439 |
| 6.00 | 7.60623 | 14.58823 | 14.53235 | 13.75 | 9.39364 | 6.31143 | 6.28480 |
| 6.25 | 7.73992 | 13.86797 | 13.81414 | 14.00 | 9.40305 | 6.22541 | 6.19921 |
| 6.50 | 7.86567 | 13.22100 | 13.16907 | 14.25 | 9.41085 | 6.14316 | 6.11735 |
| 6.75 | 7.98389 | 12.63730 | 12.58712 | 14.50 | 9.41710 | 6.06442 | 6.03901 |
| 7.00 | 8.09499 | 12.10852 | 12.05997 | 14.75 | 9.42188 | 5.98899 | 5.96396 |
| 7.25 | 8.19936 | 11.62767 | 11.58064 | 15.00 | 9.42525 | 5.91668 | 5.89201 |
| 7.50 | 8.29735 | 11.18888 | 11.14327 | 15.25 | 9.42730 | 5.84729 | 5.82298 |
| 7.75 | 8.38929 | 10.78714 | 10.74285 | 15.50 | 9.42806 | 5.78067 | 5.75669 |
| 8.00 | 8.47551 | 10.41819 | 10.37516 | 15.75 | 9.42762 | 5.71665 | 5.69301 |
| 8.25 | 8.55630 | 10.07839 | 10.03653 | 16.00 | 9.42602 | 5.65509 | 5.63177 |
| 8.50 | 8.63194 | 9.76458 | 9.72383 | 16.25 | 9.42331 | 5.59585 | 5.57285 |
| 8.75 | 8.70270 | 9.47405 | 9.43434 | 16.50 | 9.41956 | 5.53882 | 5.51612 |
| 9.00 | 8.76882 | 9.20442 | 9.16569 | 16.75 | 9.41481 | 5.48387 | 5.46147 |
| 9.25 | 8.83055 | 8.95363 | 8.91584 | 17.00 | 9.40910 | 5.43089 | 5.40878 |
| 9.50 | 8.88810 | 8.71986 | 8.68295 | 17.25 | 9.40249 | 5.37979 | 5.35795 |
| 9.75 | 8.94169 | 8.50152 | 8.46546 | 17.50 | 9.39501 | 5.33046 | 5.30890 |
| 10.00 | 8.99151 | 8.29720 | 8.26194 | 17.75 | 9.38671 | 5.28283 | 5.26153 |
| 10.25 | 9.03775 | 8.10565 | 8.07116 | 18.00 | 9.37762 | 5.23680 | 5.21576 |
| 10.50 | 9.08059 | 7.92578 | 7.89201 | 18.25 | 9.36779 | 5.19230 | 5.17152 |
| 10.75 | 9.12020 | 7.75658 | 7.72352 | 18.50 | 9.35724 | 5.14926 | 5.12872 |
| 11.00 | 9.15674 | 7.59719 | 7.56479 | 18.75 | 9.34602 | 5.10760 | 5.08731 |
| 11.25 | 9.19034 | 7.44680 | 7.41504 | 19.00 | 9.33415 | 5.06727 | 5.04721 |
| 11.50 | 9.22117 | 7.30471 | 7.27356 | 19.25 | 9.32167 | 5.02820 | 5.00837 |
| 11.75 | 9.24936 | 7.17029 | 7.13971 | 19.50 | 9.30860 | 4.99034 | 4.97073 |
| 12.00 | 9.27502 | 7.04294 | 7.01293 | 19.75 | 9.29497 | 4.95363 | 4.93424 |
| 12.25 | 9.29829 | 6.92215 | 6.89268 | 20.00 | 9.28082 | 4.91803 | 4.89885 |
| 12.50 | 9.31927 | 6.80745 | 6.77850 | | | | |

**TABLE S15:** Fine cavity-frequency scan through the $H_2$ $\sigma \rightarrow \sigma^*$ resonance ($H_2$/6-31G, $\lambda = 0.05$ a.u. along the bond, $n_{max} = 20$): exact $n_c$ and $g_c^{(2)}$, and QED-CCSD-22 $g_c^{(2)}$, at all 61 grid points. $n_c$ is exact and scaled by $10^4$; $\omega$ in eV.

## S8. Third-cumulant values

Table S16 presents the $\kappa_{3,c}$ values mentioned in the main text: the $H_2$ machine zero required by the combined-parity selection rule, the matched-basis validation against exact diagonalization for $HeH^+$ and LiH, and the aug-cc-pVDZ values for HF and $H_2O$. The signs refer to the coupling orientations of Sec. S6 for which $\lambda$ points from the heavy atom toward H for the diatomics and along $+z$ for $H_2O$, that is, toward the positive end of the charge distribution for $HeH^+$ and HF and toward the negative end for hydridic LiH and for $H_2O$. Reversing $\lambda$ maps $\hat{H}(\lambda)$ onto $\hat{H}(-\lambda)$ and flips the sign of every odd central moment exactly. The reported signs follow this reversal identity, which does not establish a universal sign relation to the permanent dipole.

For an isotropic or head–tail-symmetric distribution, opposite coupling directions have equal weight and the odd moments cancel. An individually addressed molecule retains its fixed-orientation signal if its orientation is stable during measurement. Interpreting the sign requires molecular polarity information and an optical phase reference. The third field moment $\kappa_{3,c} = \langle \delta b^\dagger \delta b \delta b \rangle$ differs from ordinary three-photon coincidence counting, which measures $\langle b^{\dagger 3} b^3 \rangle$, a third-order intensity correlation (sixth order in the field operators). Even moments are unchanged under coupling reversal and need not average to zero. General rotations change the projected dipole couplings, and geometry relaxation and nuclear motion change molecular energies and dipole matrix elements. Ensemble correlations must be constructed from the moments of the specified ensemble state and detection geometry; they do not generally equal the arithmetic average of normalized single-orientation ratios. Section S12 and Table S19 quantify the fixed-geometry orientation dependence for $H_2O$.

| system | $\kappa_{3,c}$ (CC) | $\kappa_{3,c}$ (exact) | rel. | $\acute{\kappa}_3$ (CC) |
|---|---|---|---|---|
| $H_2$/6-31G | $+3.6311\times10^{-18}$ | $+7.5563\times10^{-21}$ | $4.8\times10^{2}$ | $+0.000$ |
| $HeH^+$/STO-3G | $+2.2070\times10^{-7}$ | $+2.2081\times10^{-7}$ | $5.4\times10^{-4}$ | $+1.548$ |
| LiH/STO-3G | $-9.0735\times10^{-6}$ | $-9.1684\times10^{-6}$ | $1.0\times10^{-2}$ | $-2.955$ |
| HF/aug-cc-pVDZ | $+7.1565\times10^{-7}$ | — | — | $+0.470$ |
| $H_2O$/aug-cc-pVDZ | $-3.7225\times10^{-7}$ | — | — | $-0.133$ |

**TABLE S16:** Third photon cumulant $\kappa_{3,c} = \langle \delta\hat{b}^\dagger \delta\hat{b} \delta\hat{b} \rangle$ at $\lambda = 0.02$ a.u., $\omega = 10$ eV. $H_2$ vanishes by the combined-parity selection rule (machine zero in both methods); polar molecules give finite, signed values. $\acute{\kappa}_3 = \kappa_{3,c}/n_c^{3/2}$.

## S9. Invariant quadrature variances and Mandel parameter

The central second moments assembled from the moments of Eq. (S1) also yield the quadrature variances of the fluctuation field. With $C_{bb}=M_{\hat{b}^2}-\beta^2$ and $C_{b^\dagger b^\dagger}=M_{\hat{b}^{\dagger 2}}-\acute{\beta}^2$, the variances of $\hat{X}=(\delta\hat{b}+\delta\hat{b}^\dagger)/\sqrt{2}$ and $\hat{P}=(\delta\hat{b}-\delta\hat{b}^\dagger)/(i\sqrt{2})$ are given by

$$
\begin{aligned}
Var(X)_c &= 1/2\left(C_{bb}+C_{b^\dagger b^\dagger}+2n_c+1\right),\\
Var(P)_c &= 1/2\left(2n_c+1-C_{bb}-C_{b^\dagger b^\dagger}\right),
\end{aligned}
\tag{S9}
$$

and the Mandel parameter of the fluctuation field follows from the invariant statistics alone, $Q_c=n_c(g_c^{(2)}-1)$ for $n_c>0$. A coherent state, and hence QED-HF, gives quadrature variances of $1/2$, while the normalized Mandel parameter of its vacuum fluctuation field is undefined because $n_c=0$. Table S17 presents all five systems of the main text at $\omega\in\{5,10,20\}$ eV ($\lambda=0.02$ a.u., aug-cc-pVDZ), with exact-diagonalization rows for the two-electron systems. Across all systems $Var(X)_c$ lies above $1/2$ and $Var(P)_c$ slightly below it, so the super-Poissonian fluctuation field is weakly squeezed along $P$. These displacement-invariant length-gauge quantities complement the quadrature variances and Mandel parameters reported in the velocity gauge by Ref. 7. Comparing the same physical observable between gauges requires consistent transformation of the Hamiltonian, state, and observable; the untransformed mode expressions need not agree.

| system | method | $\omega$ (eV) | $Var(X)_c$ | $Var(P)_c$ | $Q_c$ |
|---|---|---|---|---|---|
| $H_2$ | CC | 5.0 | 0.500608 | 0.499643 | $2.338 \times 10^{-3}$ |
| $H_2$ | exact | 5.0 | 0.500608 | 0.499643 | $2.343 \times 10^{-3}$ |
| $H_2$ | CC | 10.0 | 0.500541 | 0.499774 | $1.419 \times 10^{-3}$ |
| $H_2$ | exact | 10.0 | 0.500541 | 0.499774 | $1.423 \times 10^{-3}$ |
| $H_2$ | CC | 20.0 | 0.500430 | 0.499885 | $8.577 \times 10^{-4}$ |
| $H_2$ | exact | 20.0 | 0.500430 | 0.499885 | $8.596 \times 10^{-4}$ |
| HF | CC | 5.0 | 0.500607 | 0.499584 | $3.065 \times 10^{-3}$ |
| HF | CC | 10.0 | 0.500565 | 0.499700 | $1.763 \times 10^{-3}$ |
| HF | CC | 20.0 | 0.500485 | 0.499818 | $1.030 \times 10^{-3}$ |
| $H_2O$ | CC | 5.0 | 0.500861 | 0.499443 | $3.849 \times 10^{-3}$ |
| $H_2O$ | CC | 10.0 | 0.500787 | 0.499612 | $2.245 \times 10^{-3}$ |
| $H_2O$ | CC | 20.0 | 0.500658 | 0.499773 | $1.303 \times 10^{-3}$ |
| $HeH^+$ | CC | 5.0 | 0.500157 | 0.499885 | $9.526 \times 10^{-4}$ |
| $HeH^+$ | exact | 5.0 | 0.500157 | 0.499885 | $9.528 \times 10^{-4}$ |
| $HeH^+$ | CC | 10.0 | 0.500149 | 0.499913 | $5.294 \times 10^{-4}$ |
| $HeH^+$ | exact | 10.0 | 0.500149 | 0.499913 | $5.297 \times 10^{-4}$ |
| $HeH^+$ | CC | 20.0 | 0.500132 | 0.499945 | $3.105 \times 10^{-4}$ |
| $HeH^+$ | exact | 20.0 | 0.500132 | 0.499945 | $3.107 \times 10^{-4}$ |
| LiH | CC | 5.0 | 0.501931 | 0.499279 | $7.228 \times 10^{-3}$ |
| LiH | CC | 10.0 | 0.501478 | 0.499634 | $4.118 \times 10^{-3}$ |
| LiH | CC | 20.0 | 0.501006 | 0.499844 | $2.175 \times 10^{-3}$ |

**TABLE S17:** Invariant quadrature variances and Mandel parameter of the fluctuation field **(**aug-cc-pVDZ, $\lambda = 0.02$ a.u. along the molecular/$C_{2v}$ axis; exact rows: Pauli–Fierz diagonalization, $n_{max} = 20$). A coherent state, and hence QED-HF, gives quadrature variances of $1/2$; the normalized fluctuation-field Mandel parameter is undefined at $n_c = 0$.

### S10. Software and data

All CC calculations use a locally developed spin-adapted QED-CCSD-22 code built on PySCF integrals;[8] working equations were generated with SASQ.[4] Every number in the original tables is generated programmatically from archived data files by the script that typeset these tables. The sweep drivers that produced the outputs are stored alongside them. An archival snapshot of the code and data of this Communication including source code, working equations, the exact-diagonalization reference, sweep drivers, outputs, and the table/figure generators is deposited on Zenodo (DOI: 10.5281/zenodo.22033874). We acknowledge the use of artificial intelligence in writing some of the code used in this study.

## S11. Additional LiH basis-set validation

Table S18 compares QED-CCSD-22 and QED-FCI values of $g_c^{(2)}$ for LiH with the STO-3G and 6-31G basis sets. The bond length is 1.595 Å, the cavity frequency is 10 eV, and the coupling vector lies along the bond. The STO-3G results are taken from the original validation calculations. The 6-31G calculations correlate all four electrons in 11 spatial orbitals. The electronic FCI space has 3025 determinants, and $n_{max}=20$ gives a total product-space dimension of 63,525.

The exact reference uses the same length-gauge Pauli–Fierz Hamiltonian and electronic-position dipole convention as the original validation. Both dipole self-energy contributions are retained, including the one-body term obtained from analytic quadrupole integrals. The lowest eigenpair is obtained by iterative diagonalization using the electronic FCI blocks, without storing the full electron–photon matrix. Comparisons with the dense diagonalization for $H_2$/6-31G and LiH/STO-3G at $\lambda=0.02$ a.u. and $n_{max}=8$ agree in $g_c^{(2)}$ within $1.6\times10^{-11}$ relative. The new $n_{max}=20$ eigenpair residual norms are below $8\times10^{-14}$ hartree, and the QED-CCSD-22 amplitude and de-excitation residual norms are below $10^{-12}$ in the solver convention.

The LiH/6-31G relative errors in $g_c^{(2)}$ are 0.10%, 1.76%, 9.99%, and 28.3% at $\lambda=0.005$, 0.02, 0.05, and 0.1 a.u., respectively. The large error at the upper end of the coupling sweep therefore persists in the larger basis. At $\lambda=0.02$ a.u., the error increases from 1.16% with STO-3G to 1.76% with 6-31G. These matched-basis comparisons assess the QED-CCSD-22 truncation error and do not establish convergence with respect to the electronic basis.

| Basis | $\lambda$ | $g_{c,CC}^{(2)}$ | $g_{c,FCI}^{(2)}$ | Error (%) | $\Delta E$ (hartree) | Relative cutoff change |
|---|---|---|---|---|---|---|
| STO-3G | 0.005 | 17.20224 | 17.20855 | 0.0367 | $1.05\times10^{-5}$ | $1.20\times10^{-13}$ |
| STO-3G | 0.02 | 16.94823 | 17.14727 | 1.16 | $1.05\times10^{-5}$ | $3.20\times10^{-13}$ |
| STO-3G | 0.05 | 15.64265 | 16.81290 | 6.96 | $1.11\times10^{-5}$ | $1.61\times10^{-14}$ |
| STO-3G | 0.1 | 12.28311 | 15.75840 | 22.1 | $2.69\times10^{-5}$ | $1.06\times10^{-10}$ |
| 6-31G | 0.005 | 8.62555 | 8.63420 | 0.10 | $1.14\times10^{-5}$ | $9.04\times10^{-11}$ |
| 6-31G | 0.02 | 8.48963 | 8.64188 | 1.76 | $1.14\times10^{-5}$ | $1.50\times10^{-13}$ |
| 6-31G | 0.05 | 7.80579 | 8.67216 | 9.99 | $1.26\times10^{-5}$ | $9.13\times10^{-13}$ |
| 6-31G | 0.1 | 6.21656 | 8.66921 | 28.3 | $5.84\times10^{-5}$ | $8.65\times10^{-10}$ |

TABLE S18: Additional LiH basis-set validation at $\omega=10$ eV. The error in $g_c^{(2)}$ is $100|g_{c,CC}^{(2)}/g_{c,FCI}^{(2)}-1|$. The energy difference is $\Delta E=E_{CC}-E_{FCI}$. The last column gives the relative change in the exact $g_c^{(2)}$ when increasing $n_{max}$ from 12 to 20. Couplings are in atomic units, and energies are in Hartree.

## S12. Fixed-geometry orientation dependence

Table S19 compares three orthogonal cavity polarizations for $H_2O$/aug-cc-pVDZ at $\omega=10$ eV and $|\lambda|=0.02$ a.u. The geometry is identical to Sec. S6, with the molecule in the $yz$ plane and its $C_2$ axis along $z$. All ten electrons are correlated using QED-CCSD-22 with the full dipole self-energy. The amplitude and de-excitation residual norms are below $10^{-12}$. The $z$ result reproduces Table S9.

Both $n_c$ and $G_c^{(2)}$ remain positive for all three polarizations. The normalized coherence remains super-Poissonian and varies with orientation. For the two perpendicular polarizations, reflection of the corresponding molecular coordinate combined with photon parity forces the odd central moments to vanish. The computed $|\kappa_{3,c}/n_c^{3/2}|$ is below $10^{-6}$ in both cases. These fixed-geometry calculations do not give an isotropic ensemble average or include nuclear relaxation.

| Polarization | $n_c$ | $G_c^{(2)}$ | $g_c^{(2)}$ | $\kappa_{3,c}$ |
|---|---|---|---|---|
| $z$ | $1.99067\times10^{-4}$ | $4.86441\times10^{-7}$ | 12.27529 | $-3.72247\times10^{-7}$ |
| $y$ | $2.20716\times10^{-4}$ | $5.59985\times10^{-7}$ | 11.49498 | $8.17047\times10^{-21}$ |
| $x$ | $1.89041\times10^{-4}$ | $4.47763\times10^{-7}$ | 12.52957 | $2.55796\times10^{-21}$ |

TABLE S19: Orientation dependence of the $H_2O$ photon moments at $\omega=10$ eV and $|\lambda|=0.02$ a.u. (QED-CCSD-22/aug-cc-pVDZ). Polarizations refer to the coordinates of Sec. S6. The fourth central moment is $G_c^{(2)}=\langle \delta b^\dagger \delta b^\dagger \delta b \delta b \rangle$, with $g_c^{(2)}=G_c^{(2)}/n_c^2$. The third cumulant is $\kappa_{3,c}=\langle \delta b^\dagger \delta b \delta b \rangle$.